\documentclass[11pt]{article}
\pdfoutput=1
\usepackage{amssymb}
\usepackage{amsmath}
\usepackage{amstext}
\usepackage{graphicx,epsfig}
\usepackage{epsfig}
\usepackage{verbatim} 
\usepackage{fancybox}
\usepackage{color}
\usepackage{ulem}
\usepackage{enumitem}
\usepackage{subfigure}
\usepackage{bbm}
\usepackage{parskip}
\usepackage[numbers,sort&compress]{natbib}
\usepackage{ytableau}

\newcommand*{\medotimes}{\raisebox{-0.3ex}{\scalebox{1.5}{$\otimes$}}}
\newcommand*{\medoplus}{\raisebox{-0.3ex}{\scalebox{1.5}{$\oplus$}}}

\newcommand{\Comment}[1]{{}}
\definecolor{darkblue}{rgb}{0.15,0.35,0.55}
\definecolor{reddish}{rgb}{0.65, 0.2, 0.2}
\usepackage[linktocpage=true]{hyperref}
\hypersetup{
colorlinks=true,
citecolor=darkblue,
linkcolor=reddish,
urlcolor=darkblue,
pdfauthor={},
pdftitle={},
pdfsubject={}
}

\newcommand{\be}{\begin{equation}}
\newcommand{\ee}{\end{equation}}
\newcommand{\bea}{\begin{eqnarray}}
\newcommand{\eea}{\end{eqnarray}}
\newcommand{\beas}{\begin{eqnarray*}}
\newcommand{\eeas}{\end{eqnarray*}}
\newcommand{\nn}{\nonumber}

\def\({\left(}
\def\){\right)}

\newcommand{\rd}{{\rm d}}

\def\gsim{ \lower .75ex \hbox{$\sim$} \llap{\raise .27ex \hbox{$>$}} }
\def\lsim{ \lower .75ex \hbox{$\sim$} \llap{\raise .27ex \hbox{$<$}} }

\def\xyma{\xymatrix@M.7em}
\def\xymas{\xymatrix@M.1em}
\newcommand{\ba}{\begin{eqnarray}}
\newcommand{\ea}{\end{eqnarray}}

\title{}
\author{}

\numberwithin{equation}{section}

\begin{document}
%
\renewcommand{\thefootnote}{\fnsymbol{footnote}}
~
\vspace{2.5truecm}
\begin{center}
{\LARGE \bf{Manifest Duality for Partially Massless Higher Spins}}
\end{center} 

\vspace{1truecm}
\thispagestyle{empty}
\centerline{{\Large Kurt Hinterbichler${}^{\rm a,}$\footnote{\href{mailto:khinterbichler@perimeterinstitute.ca}{\texttt{kurt.hinterbichler@case.edu}}} and Austin Joyce${}^{\rm b,}$\footnote{\href{mailto:ajoy@uchicago.edu}{\texttt{ajoy@uchicago.edu}}}}}
\vspace{.7cm}

\centerline{{\it ${}^{\rm a}$CERCA, Department of Physics,}}
 \centerline{{\it Case Western Reserve University, 10900 Euclid Ave, Cleveland, OH 44106}} 
 \vspace{.5cm}

\centerline{\it ${}^{\rm b}$Enrico Fermi Institute and Kavli Institute for Cosmological Physics,}
\centerline{\it University of Chicago, Chicago, IL 60637}

 \vspace{.8cm}
\begin{abstract}
\noindent
In four dimensions, partially massless fields of all spins and depths possess a duality invariance akin to electric-magnetic duality.
We construct metric-like gauge invariant curvature tensors for partially massless fields of all integer spins and depths, and show how the partially massless equations of motion can be recovered from first order field equations and Bianchi identities for these curvatures. This formulation displays duality in its manifestly local and covariant form, in which it acts to interchange the field equations and Bianchi identities.
\end{abstract}

\newpage

\setcounter{tocdepth}{2}
\tableofcontents
\newpage
\renewcommand*{\thefootnote}{\arabic{footnote}}
\setcounter{footnote}{0}

\section{Introduction and summary}
On flat space, spin-$s$ fields fall into a binary classification; they are either massive or massless. On curved backgrounds, there is a more intricate structure. The (anti) de Sitter group possesses exotic irreducible representations which do not have any flat space analogues. These ``partially massless" (PM) fields come in various depths, labelled by $t\in \left\{0,1,\ldots,s-1\right\}$, and display gauge invariances which remove helicity components $0,1,\ldots, t$ from the massive field, leaving a number of degrees of freedom intermediate between that of a massless and a massive field~\cite{Deser:1983tm,Deser:1983mm,Higuchi:1986py,Brink:2000ag,Deser:2001pe,Deser:2001us,Deser:2001wx,Deser:2001xr,Zinoviev:2001dt,Skvortsov:2006at,Skvortsov:2009zu}.

Partially massless fields have recently seen renewed interest due to possible connections between a PM spin-2 field and cosmology (see e.g.,~\cite{deRham:2013wv} and the review \cite{Schmidt-May:2015vnx}).  There have been attempts and no-go's bearing on the construction of a self-interacting theory of a partially massless spin-2~\cite{Deser:2013uy,deRham:2013wv,Joung:2014aba,Zinoviev:2014zka,Alexandrov:2014oda}, and extensive exploration of the properties of the linear theory and other possible nonlinear extensions~\cite{Zinoviev:2006im,Hassan:2012gz,Hassan:2012rq,Hassan:2013pca,Zinoviev:2014zka,Garcia-Saenz:2014cwa,Hinterbichler:2014xga,Hassan:2015tba,Hinterbichler:2015nua,Cherney:2015jxp,Gwak:2015vfb,Gwak:2015jdo,Garcia-Saenz:2015mqi}. 

It was shown in~\cite{Deser:2013xb,Deser:2014ssa}, using a non-manifestly covariant 3+1 formulation, that partially massless fields in four dimensions possess a duality invariance in a manner akin to electric-magnetic duality \cite{Deser:1976iy}.\footnote{Similarly, it is known that massless and massive spin-$s$ fields on various backgrounds possess such a duality invariance~\cite{Hull:2001iu,Francia:2002aa,Francia:2002pt,Bekaert:2003az,Deser:2004xt,Henneaux:2004jw,Francia:2005bu,Julia:2005ze,Julia:2005wg,Francia:2007ee,Leigh:2007wf,Bunster:2012km,Henneaux:2015cda}, which extends to fields of arbitrary mixed symmetry~\cite{Bekaert:2002dt,deMedeiros:2002qpr,Bekaert:2015fwa}.}  Electromagnetic duality, since its origins almost a century ago \cite{Dirac:1931kp,Dirac:1948um}, has played a central role in many of the advances of modern theoretical physics (see e.g. the reviews \cite{Harvey:1996ur,AlvarezGaume:1997ix,Obers:1998fb}), so it is naturally of interest to explore its implications in the partially massless case.

Our goal will be to see the duality of the partially massless fields in its manifestly covariant form.   In general, accomplishing this requires casting the field equations into a first order form, which is different from the standard, second order, Fronsdal approach~\cite{Fronsdal:1978rb,Fang:1978wz}. In particular, the equations of motion are reproduced by taking as the fundamental object a gauge-invariant curvature.  For massless spin-1 and spin-2, these are just the standard Maxwell field strength and Riemann curvatures respectively, but for massless higher spins it requires the introduction of new generalized curvatures~\cite{deWit:1979pe,Damour:1987vm}.  (For some reviews of various aspects of higher spin theory, see e.g.~\cite{Sorokin:2004ie,Bouatta:2004kk,Fotopoulos:2008ka,Bekaert:2010hw,Sagnotti:2013bha,Rahman:2013sta,Rahman:2015pzl}.)  In~\cite{Hinterbichler:2014xga}, the duality of~\cite{Deser:2013xb} was displayed in covariant form, with manifest de Sitter invariance, for the partially massless spin-2. Here we generalize this construction to partially massless fields of arbitrary integer spin and depth.

The duality-covariant equations for a spin-$s$ depth-$t$ field will be formulated in terms of an $(s+t+1)$-index tensor with the symmetry type
\be
\ytableausetup{centertableaux,boxsize=1.4em}
{\cal K}~\in~\begin{ytableau}
 ~\\~
\end{ytableau}
~~\medotimes~~\begin{array}{|c c c c c|}\hline
&s-1\!\!\!\!\!\!&\!\!\!&&\\
\hline
\!\!\!  &\!\! t~~~  \vline\!\!\!\!\!\!\\
\cline{1-2}
\end{array}~. \label{tablequxinto}
\ee
This tensor is then constrained to satisfy the following Maxwell-like equations,
\begin{align}
{\rm tr}\ast{\cal K} &= 0\ ,~~~~~~~~~~~~~~~\rd{\cal K} = 0~,\label{introeoms1} \\ 
{\rm tr}\,{\cal K} &= 0\ ,~~~~~~~~~~~~\rd\ast{\cal K} = 0~, \label{introeoms2}
\end{align}
where the exterior derivative and Hodge star act covariantly on the first tensor factor of \eqref{tablequxinto}.
The equations \eqref{introeoms1} are Bianchi-like identities and \eqref{introeoms2} are dynamical equations which together will reproduce the equations of motion of the partially massless field.  

The algebraic Bianchi identity restricts the form of the tensor \eqref{tablequxinto}, projecting away many of the components and leaving a residual tensor of the symmetry type~
\be
\scalebox{1}{$
\begin{array}{|c c c c c|}\hline
&s\!\!\!\!\!\!&\!\!\!&&\\
\hline
\!\!\!  &\!\!\! t+1~~  \vline\!\!\!\\
\cline{1-2}
\end{array}
$}\,,
\ee
which will become the gauge invariant PM curvature.\footnote{Note that this tensor has the same symmetry type as the frame-like curvature tensors of Skvortsov and Vasiliev~\cite{Skvortsov:2006at}. Here we provide an alternative metric-like construction of these tensors and show how to reproduce the on-shell equations of motion for a PM  field from these curvatures.}  The second of the Bianchi identities~\eqref{introeoms2} is a differential Bianchi identity, and fits into a differential complex of the form
\be
\begin{array}{|c c c c c|}\hline
&\!\!\!&t&&\!\!\!\\
\hline
\end{array}
\xrightarrow{\hspace*{.75cm}}~
\begin{array}{|c c c c c|}\hline
&~&s&~&\\
\hline
\end{array}~
\xrightarrow{\hspace*{.75cm}}~
\begin{array}{|c c c c c|}\hline
&s\!\!\!\!\!\!&\!\!\!&&\\
\hline
\!\!\!  &\!\!\! t+1~~  \vline\!\!\!\\
\cline{1-2}
\end{array}~
\xrightarrow{\hspace*{.75cm}}~
\begin{array}{|c c c c c|}\hline
&s\!\!\!\!\!\!&\!\!\!&&\\
\hline
\!\!\!  &\!\!\! t+1~~  \vline\!\!\!\\
\cline{1-2}
~~~\!\!~\vline\!\!\!\\
\cline{1-1}
\end{array}
\xrightarrow{\hspace*{.75cm}}~
\cdots \label{gencomplexintro}
\ee
We will use the assumption of trivial cohomology of this complex to write the tensor ${\cal K}$  as an appropriately symmetrized $(t+1)^{\rm th}$ derivative of a totally symmetric rank-$s$ gauge potential. 

We will then turn to the equations \eqref{introeoms2}, from which we will recover the on shell equations of motion for the partially massless field.  For the higher depths, this involves generalizing the approach of~\cite{Francia:2002aa,Francia:2002pt} to the partially massless (A)dS setting.
From this formulation, it is manifest that the equations \eqref{introeoms1}, \eqref{introeoms2} are invariant in $D=4$ under the duality rotation $\delta {\cal K} = \ast{\cal K}$, which is the PM analogue of electric-magnetic duality.

The curvature construction for depth-$t$ fields will be strongly reminiscent of that of a spin-$(t+1)$ massless field's.  This will reinforce the notion that fields of different spins but the same depth of partial masslessness have more in common with each other than do fields which have different depths but the same spin.  The arguments for the lower depths $t=0,1,$ are somewhat different from those for the higher depths $t\geq 2$, so we will treat them separately, organizing the discussion according to the depth of partial masslessness. 

We begin in section~\ref{sec:introtoPM} by reviewing some salient features of partially massless fields, including the on-shell equations of motion that we aim to reproduce. We then consider partially massless fields of depth $t=0$ in section~\ref{sec:maximaldepth} and show how their equations of motion can be reproduced by considering a generalized Maxwell tensor. This is a more-or-less direct generalization of the story for the PM spin-2 case presented in~\cite{Hinterbichler:2014xga} (see also~\cite{Cherney:2015jxp}).  We next consider depth $t=1$ fields and perform a similar construction in section~\ref{sec:deptht1}.  Here the construction follows the pattern of linearized Einstein gravity.  Finally, we discuss the case of depths $t\geq 2$ in section~\ref{sec:higherdepth}. The main difference in this case is that the curvature tensor has $\geq 3$ derivatives, so the second order equations of motion are recovered in a somewhat subtle way, similar to the massless case for $s \geq 3$~\cite{Francia:2002aa,Francia:2002pt,Bekaert:2003az}. In Appendix~\ref{spin3workedout}, we  work out the equations of motion for both partially massless points of a spin-3 field on de Sitter space from the off-shell Lagrangian starting point. This is provided for  convenience to illustrate the relationship between this standard viewpoint and the formalism which we adopt in the rest of the paper. We comment on some natural future directions in section~\ref{sec:conclusions}.

\vspace{-.5cm}
\paragraph{Conventions:}We use the mostly plus metric signature. We (anti) symmetrize tensors with unit weight, e.g., $S_{(\mu\nu)} = \frac{1}{2}(S_{\mu\nu}+S_{\nu\mu})$. We work on de Sitter space of dimension $D$ and Hubble radius $1/H$ throughout.  The curvature tensors of this de Sitter space are given by
\begin{equation*}
R_{\mu\nu\rho\sigma} = H^2\left(g_{\mu\rho}g_{\nu\sigma} - g_{\mu\sigma}g_{\nu\rho}\right),~~~~~~~
R_{\mu\nu} = (D-1)H^2 g_{\mu\nu}\,,~~~~~~~
R = D(D-1)H^2.
\end{equation*}
All of our formulae apply equally well to anti-de Sitter by taking $H^2 \mapsto -L^{-2}$, with $L$ the AdS radius. We define the depth, $t$, of a partially massless field to be the highest helicity component removed by a gauge symmetry or, equivalently, the number of indices on the gauge parameter.\footnote{Note that this definition of the depth differs from some papers in the literature, which define the depth by the number of derivatives in the gauge transformation; it is straightforward to convert between these conventions by sending $t\mapsto s-t$.} Young tableaux are employed in the manifestly antisymmetric convention, and on the tensors we use commas to delineate anti-symmetric groups of indices corresponding to columns of length two or greater.  The projector onto a tableau with row lengths $r_1, r_2,\cdots$ is denoted $P_{r_1,r_2,\cdots}$ where the indices to be projected should be obvious from context.  The action of the projector is to first symmetrize the indices in each row, and then anti-symmetrize the indices in each column, with an overall normalization fixed so that $P_{r_1,r_2,\cdots}^2=P_{r_1,r_2,\cdots}$.  An excellent introduction to Young tableaux can be found in section 4 of~\cite{Bekaert:2006py} or the book \cite{Tung:1985na}.

\section{Partially massless equations and complexes}
\label{sec:introtoPM}
A spin-$s$ field of mass $m$ on ${\rm (A)dS}_D$ is carried by a totally symmetric tensor $\ell_{\mu_1\cdots\mu_s}$ which obeys the on-shell equations of motion
\be
\left(\square-H^2\left[D+(s-2)-(s-1)(s+D-4)\right]-m^2\right)\ell_{\mu_1\cdots\mu_s} = 0\,,~~~~\nabla^\nu\ell_{\nu\mu_2\cdots\mu_s} = 0\,,~~~~\ell^\nu_{\ \nu\mu_3\cdots\mu_{s}} = 0.
\label{massivefields}
\ee
At generic values of the mass, these equations propagate
\be \frac{(D-3+2 s) (D-4+s)!}{s!(D-3)!} \ee
degrees of freedom.

\subsection{Partially massless points}

These massive fields, at particular values of the mass, can develop a gauge invariance which removes a subset of the helicity components of the representation. 
A spin-$s$ field has $(s-1)$ partially massless points, labeled by the depth $t\in \left\{0,1,\ldots, s-1\right\}$~\cite{Deser:2001xr,Zinoviev:2001dt}, which occur at the masses
\be
m^2 = (s-t-1)(s+t+D-4)H^2\,.
\label{pmpoints}
\ee
The value $t = s-1$ corresponds to the massless theory.
At these special values of the mass, a depth-$t$ partially massless field possesses a gauge invariance with a $t$-index totally symmetric gauge parameter, which removes the components of the massive field with helicity $\leq t$. Combining~\eqref{massivefields} and~\eqref{pmpoints}, the on-shell equations for a partially massless field of spin-$s$ and depth-$t$ are~\cite{Deser:2001xr,Zinoviev:2001dt,Hallowell:2005np},
\be
\left(\square-H^2\left[D+(s-2)-t(D+t-3)\right]\right)\ell_{\mu_1\cdots\mu_s} = 0~,~~~~~~\nabla^\nu\ell_{\nu\mu_2\cdots\mu_s} = 0~,~~~~~~\ell^\nu_{\ \nu\mu_3\cdots\mu_{s}} = 0,
\label{spinsdepthteom}
\ee
which has a gauge invariance
\begin{align}
 \delta \ell_{\mu_1\cdots\mu_s} &= \nabla_{(\mu_{t+1}}\nabla_{\mu_{t+2}}\cdots\nabla_{\mu_{s}}\xi_{\mu_{1}\cdots\mu_{t})}+\cdots \label{pmintogt}  \\ \nn
\ \ \ \ &= 
\left\{\begin{array}{l}
P_{s}\left(\prod_{n=1}^{\frac{s-t}{2}}\left[\nabla_{\mu_n}\nabla_{\mu_{n+\frac{s-t}{2}}}+(2n-1)^2H^2g_{\mu_n\mu_{n+\frac{s-t}{2}}}\right]\right)\xi_{\mu_{s-t+1}\cdots\mu_s}~~~~~~~~~~\,{\rm for}~(s-t)~{\rm even}\ \ \vspace{.15cm}\\
P_{s}\left(\prod_{n=1}^{\frac{s-t-1}{2}}\left[\nabla_{\mu_n}\nabla_{\mu_{n+\frac{s-t-1}{2}}}+(2n)^2H^2g_{\mu_n\mu_{n+\frac{s-t-1}{2}}}\right]\right)\nabla_{\mu_{s-t}}\xi_{\mu_{s-t+1}\cdots\mu_s}~{\rm for}~(s-t)~{\rm odd}. \ \ 
\end{array}\right. \ \ 
\end{align}
Here the ellipses stand for ${\cal O}(H^2)$ terms with fewer derivatives, which we can write in the indicated factorized form, with $P_s$ a projector onto the totally symmetric $s$-index part.

The gauge parameter, $\xi_{\mu_{1}\cdots\mu_{t}}$, is a totally symmetric tensor which is itself restricted to satisfy the on-shell equations
\be 
\Big(\square  + H^2\left[(s-1)(D+s-2)-t\right]\Big)\xi_{\mu_{1}\cdots\mu_{t}}=0,\ \ ~~~~~ \nabla^{\nu}\xi_{\nu\mu_{2}\cdots\mu_{t}}=0,\ \ \ ~~~~~  \xi^\nu_{\ \nu\mu_{3}\cdots\mu_{t}} =0,\label{genresga}\ee
so that the equations~\eqref{spinsdepthteom} are on-shell gauge invariant.  The values of the mass in \eqref{pmpoints}, as well as the form of the ${\cal O}(H^2)$ terms in \eqref{pmintogt} and \eqref{genresga}, are completely fixed by requiring that the system \eqref{massivefields} have the partially massless symmetry with the leading derivative part $\delta \ell_{\mu_1\cdots\mu_s} = \nabla_{(\mu_{t+1}}\nabla_{\mu_{t+2}}\cdots\nabla_{\mu_{s}}\xi_{\mu_{1}\cdots\mu_{t})}$.

A partially massless field of spin-$s$ and depth-$t$ possesses helicity components $\{\pm(t+1),\cdots,\pm s\}$ and propagates
\be
 \frac{\left(D-3+2s\right)\left(D-4+s\right)!}{s!(D-3)!} - \frac{\left(D-3+2t\right)\left(D-4+t\right)!}{t!(D-3)!}
\ee
physical degrees of freedom in $D$ spacetime dimensions.  These degrees of freedom transform irreducibly under an exotic representation of the (anti) de Sitter group which has no true flat-space counterpart.\footnote{In the flat space limit ($H\to0$) the partially massless representation becomes reducible and breaks up into a sum of flat space massless helicity representations $\{\pm(t+1),\cdots,\pm s\}$.}

\subsection{Differential complex\label{diffcomplexsect}}

Central to the arguments that follow will be the following sequence of first order differential operators mentioned in the introduction,
\be
\begin{array}{|c c c c c|}\hline
&\!\!\!&t&&\!\!\!\\
\hline
\end{array}
\xrightarrow{
\overset{\rd^{(s,t)}_1}{\hspace*{.75cm}}}~
\begin{array}{|c c c c c|}\hline
&~&s&~&\\
\hline
\end{array}~
\xrightarrow{
\overset{\rd^{(s,t)}_2}{\hspace*{.75cm}}}~
\begin{array}{|c c c c c|}\hline
&s\!\!\!\!\!\!&\!\!\!&&\\
\hline
\!\!\!  &\!\!\! t+1~~  \vline\!\!\!\\
\cline{1-2}
\end{array}~
\xrightarrow{
\overset{\rd^{(s,t)}_3}{\hspace*{.75cm}}}~
\begin{array}{|c c c c c|}\hline
&s\!\!\!\!\!\!&\!\!\!&&\\
\hline
\!\!\!  &\!\!\! t+1~~  \vline\!\!\!\\
\cline{1-2}
~~~\!\!~\vline\!\!\!\\
\cline{1-1}
\end{array}
\xrightarrow{\hspace*{.75cm}}~
\cdots
\label{depthtspinscomplex}
\ee
where the $\rd^{(s,t)}$ operators act as
\bea
\label{gendc1} 
&& \left(\rd^{(s,t)}_1 \xi\right)_{\mu_1\cdots \mu_s}\propto \nabla_{(\mu_{t+1}}\nabla_{\mu_{t+2}}\cdots\nabla_{\mu_{s}}\xi_{\mu_{1}\cdots\mu_{t})}+\cdots\\  \nn
&&\ \ \ \ \ \ \ \ \ \propto
\left\{\begin{array}{l}
P_{s}\left(\prod_{n=1}^{\frac{s-t}{2}}\left[\nabla_{\mu_n}\nabla_{\mu_{n+\frac{s-t}{2}}}+(2n-1)^2H^2g_{\mu_n\mu_{n+\frac{s-t}{2}}}\right]\right)\xi_{\mu_{s-t+1}\cdots\mu_s}~~~~~~~~~~~{\rm for}~(s-t)~{\rm even}\vspace{.15cm} \\
P_{s}\left(\prod_{n=1}^{\frac{s-t-1}{2}}\left[\nabla_{\mu_n}\nabla_{\mu_{n+\frac{s-t-1}{2}}}+(2n)^2H^2g_{\mu_n\mu_{n+\frac{s-t-1}{2}}}\right]\right)\nabla_{\mu_{s-t}}\xi_{\mu_{s-t+1}\cdots\mu_s}~{\rm for}~(s-t)~{\rm odd}
\end{array}\right.,  \\
&& \left(\rd^{(s,t)}_2\ell\right)_{\mu_1\nu_1,\cdots,\mu_{t+1}\nu_{t+1},\mu_{t+2}\cdots\mu_{s}} \propto P_{s,t+1}\nabla_{\nu_{1}}\cdots\nabla_{\nu_{t+1}}\ell_{\mu_{1}\cdots\mu_{s}}+\cdots \label{gendc2}\\ \nn
&&\ \ \ \ \ \ \ \ \  \propto
\left\{\begin{array}{l}P_{s,t+1}\left(\prod_{n=1}^{\frac{t+1}{2}}\left[\nabla_{\nu_n}\nabla_{\nu_{n+\frac{t+1}{2}}}+(2n-1)^2H^2g_{\nu_n\nu_{n+\frac{t+1}{2}}}\right]\right)\ell_{\mu_1\cdots\mu_s}~~{\rm for}~t~{\rm odd}\vspace{.15cm}\\
P_{s,t+1}\left(\prod_{n=1}^{\frac{t}{2}}\left[\nabla_{\nu_n}\nabla_{\nu_{n+\frac{t}{2}}}+(2n)^2H^2g_{\nu_n\nu_{n+\frac{t}{2}}}\right)\right]\nabla_{\nu_{t+1}}\ell_{\mu_1\cdots\mu_s}~~~~~{\rm for}~t~{\rm even}
\end{array} \right.\,, \\
&& \left(\rd^{(s,t)}_3{\cal K}\right)_{\mu_1\nu_1\rho,\mu_2\nu_2,\cdots,\mu_{t+1}\nu_{t+1},\mu_{t+2}\cdots\mu_{s}} \propto  P_{s,t+1,1} \nabla_{\rho} {\cal K}_{\mu_1\nu_1,\cdots,\mu_{t+1}\nu_{t+1},\mu_{t+2}\cdots\mu_{s}}, \label{gendc3} \\
&&~~~~~~~~~~\!~\vdots \nn
\eea
Here the $P$ are projectors onto the Young tableaux that appear in~\eqref{depthtspinscomplex},
and the ellipses are lower derivative terms proportional to $H^{2}$ which can be written in the indicated factorized form. 

The key property of this sequence of operators is nilpotency,
\be \rd^{(s,t)}_{i+1}\circ\rd^{(s,t)}_{i}=0,\ \ \ \ \ \ \ i=1,2,\cdots, \label{d2is0}\ee
which makes it into a differential complex.  The ${\cal O}\left(H^2\right)$ terms in the $\rd^{(s,t)}$ operators are uniquely fixed by the requirement \eqref{d2is0}.\footnote{We may gain more insight into the form of these operators by considering embedding space.  Given a $(D+1)$-dimensional Minkowski space with coordinates $X^A$ and metric, $\eta_{AB}={\rm diag}\{-1,1,1,\cdots\}$, ${\rm dS}_D$ is realized as the surface $\eta_{AB}X^AX^B= H^{-2}$ (the $\rm{AdS}_D$ case follows similarly, only with a two-time embedding space).  Letting $e_\mu^{\ A}={dX^A\over d x^\mu}$ be the projectors onto the ${\rm dS}_D$ with intrinsic coordinates $x^\mu$, we can assign to each $D$-dimensional tensor, $T_{\mu_1\cdots \mu_r}$, on ${\rm dS}_D$ (here $T$ is either the gauge parameter, gauge field or gauge field strength, with $r$ the number of indices) a corresponding $(D+1)$-dimensional tensor, $\tilde T_{A_{1}\cdots A_{r}}$, in the embedding space, which satisfies a homogeneity and scaling condition
\be X^A\partial_A \tilde T_{A_{1}\cdots A_{r}}=\left(s-1+t\right)T_{A_{1}\cdots A_{r}},\ \ \ \ X^A\tilde T_{AA_{2}\cdots A_{r}}=0.\ee
The ${\rm dS}_D$ tensor is then recovered from the embedding space tensor by pulling back to the ${\rm dS}_D$ surface,
\be T_{\mu_1\cdots \mu_r}=\sqrt{X^2} ^{-(s-1-t)}e_{\mu_1}^{\ A_1}\cdots e_{\mu_r}^{\ A_r}\ \tilde T_{A_{1}\cdots A_{r}}.\ee
The expressions \eqref{gendc1}, \eqref{gendc2}, \eqref{gendc3} descend from simple Young projections of flat derivatives,
\bea
&& \left(\rd^{(s,t)}_1 \tilde \xi\right)_{A_1\cdots A_s}\propto \partial_{(A_{t+1}}\partial_{A_{t+2}}\cdots\partial_{A_{s}}\tilde\xi_{A_{1}\cdots A_{t})}\, , \nn\\ 
&& \left(\rd^{(s,t)}_2\tilde\ell\right)_{A_1B_1,\cdots,A_{t+1}B_{t+1},A_{t+2}\cdots A_{s}} \propto P_{s,t+1}\partial_{B_{1}}\cdots\partial_{B_{t+1}}\tilde\ell_{A_{1}\cdots A_{s}}\, ,\\ \nn
&& \left(\rd^{(s,t)}_3\tilde{\cal K}\right)_{A_1B_1 C ,A_2B_2,\cdots,A_{t+1}B_{t+1},A_{t+2}\cdots A_{s}} \propto  P_{s,t+1,1} \partial_{ C } \tilde{\cal K}_{A_1B_1,\cdots,A_{t+1}B_{t+1},A_{t+2}\cdots A_{s}}\, , \\
&&~~~~~~~~~~\!~\vdots \nn
\eea
The property \eqref{d2is0} is now manifest, and so it must reproduce the intrinsic ${\rm dS}_D$ expressions upon reduction.  See \cite{Bekaert:2010hk,Joung:2012rv,Joung:2012hz,Bekaert:2013zya} for more on the embedding space formulation of partially massless fields.
}

This complex generalizes the de Rham complex on (A)dS (which appears as the special case $s=1$, $t=0$)
and contains, from the left to right respectively, the gauge parameters, gauge fields, field strengths and Bianchi identities of the partially massless spin-$s$ field of depth $t$. It generalizes to ${\rm (A)dS}_D$ and to higher spin the complexes of \cite{DuboisViolette:2001jk,Bekaert:2002dt,Hinterbichler:2014xga}.  In particular, \eqref{d2is0} implies that if the curvature tensor is written in terms of a gauge field as ${\cal K}=\rd^{(s,t)}_2\ell$, then it is gauge invariant under $\delta \ell=\rd^{(s,t)}_1\xi$. 

We will assume that the cohomology of this complex is trivial.\footnote{For appropriate boundary conditions on a patch of trivial topology it should be possible to prove trivial cohomology in a manner similar to~\cite{pjoliver:1982,DuboisViolette:1999rd,DuboisViolette:2001jk,Bekaert:2002dt}.  It would also be interesting to investigate the consequences of a non-trivial cohomology due to the presence of non-trivial boundaries, topologies, or PM monopoles such as those of \cite{Hinterbichler:2015nua}.}  This implies, for example, that if ${\cal K}$ is annihilated by the operator $\rd^{(s,t)}_3$, it may be written in terms of a spin-$s$ potential $\ell$, and so we have a two way implication,
\be
\rd^{(s,t)}_3 {\cal K}=0 \iff {\cal K}=\rd^{(s,t)}_2 \ell.
\ee

\subsection{Field strength and equation of motion\label{fieldseomsubsec}}

The field strength of a spin-$s$ depth-$t$ partially massless field will start out as an $(s+t+1)$-index tensor ${\cal K}$ with the following symmetry type,
\be
{\cal K}_{\mu_1\nu_1|\mu_2\nu_2,\cdots,\mu_{t+1}\nu_{t+1},\alpha_{1}\cdots\alpha_{s-1-t}}\in~\ytableausetup{centertableaux,boxsize=1.4em}
\begin{ytableau}
 ~\\~
\end{ytableau}
~~\medotimes~~\begin{array}{|c c c c c|}\hline
&s-1\!\!\!\!\!\!&\!\!\!&&\\
\hline
\!\!\!  &\!\! t~~~  \vline\!\!\!\!\!\!\\
\cline{1-2}
\end{array}~.
\label{spinsdepthtint}
\ee
It is anti-symmetric in its first two indices, and in its remaining indices it has the symmetry of the two-row Young diagram with rows of length $s-1$ and $t$.  It has no symmetries among the first two indices and the rest, and no constraints on traces.

We will show that under the assumption of trivial cohomology for the complex of section \ref{diffcomplexsect}, the equations of motion \eqref{spinsdepthteom} for a depth-$t$ PM field of spin-$s$ are equivalent to the following Maxwell-like set of equations for the tensor ${\cal K}$,
\begin{align}
\label{pmssbianchi}
{\rm tr}\ast {\cal K} & = 0\, ,~~~~~~~~~~~~~~~~\rd\, {\cal K} =  0\, ,\\
{\rm tr}\, {\cal K} & = 0\, ,~~~~~~~~~~~~~\rd\ast {\cal K} = 0~.
\label{pmsseoms}
\end{align}
Here the Hodge star and exterior $\rd$ operator act only with respect to the first set of indices,\footnote{However, it is to be understood that the Christoffel symbols associated with the covariant derivative are to included for all the indices, i.e., the equations are ${\rm (A)dS}$ covariant.}
\begin{align}
\left(\ast {\cal K}\right)_{\mu_1\nu_1\beta_1\cdots\beta_{D-2}|\mu_2\nu_2,\cdots,\mu_{t+1}\nu_{t+1},\alpha_{1}\cdots\alpha_{s-1-t}}  &= {1\over 2}\epsilon_{\mu_1\nu_1\beta_1\cdots\beta_{D-2}}^{~~~~~~~~~~~~~~~\rho\sigma}{\cal K}_{\rho\sigma|\mu_2\nu_2,\cdots,\mu_{t+1}\nu_{t+1},\alpha_{1}\cdots\alpha_{s-1-t}},\\
\left(\rd {\cal K}\right)_{\rho\mu_1\nu_1|\mu_2\nu_2,\cdots,\mu_{t+1}\nu_{t+1},\alpha_{1}\cdots\alpha_{s-1-t}}&=3\nabla_{[\rho}{\cal K}_{\mu_1\nu_1]|\mu_2\nu_2,\cdots,\mu_{t+1}\nu_{t+1},\alpha_{1}\cdots\alpha_{s-1-t}},
\end{align}
and the trace is between one index in the first set and the first index in the second set,
\be \left( {\rm tr}\,{\cal K}\right)_{~\nu_1|\nu_2,\cdots,\mu_{t+1}\nu_{t+1},\alpha_{1}\cdots\alpha_{s-1-t}} = {\cal K}^\rho_{~\nu_1|\rho\nu_2,\cdots,\mu_{t+1}\nu_{t+1},\alpha_{1}\cdots\alpha_{s-1-t}}. \ee

The equations~\eqref{pmssbianchi} consist of two Bianchi-type identities; the first is an algebraic Bianchi identity, which will restrict the symmetry type of the tensor ${\cal K}$, while the second is a differential Bianchi identity which will tell us that ${\cal K}$ can be written as the field strength of a gauge field.   The equations~\eqref{pmsseoms} will then become field equations which reproduce the equations of motion for the spin-$s$ field.  This is the generalization of the story for electromagnetism or for PM spin-2~\cite{Hinterbichler:2014xga}.

\subsection{Duality}

In the case $D=4$, $\ast {\cal K}$ and ${\cal K}$ have the same number of indices and carry the same representation.  In this case, the  equations \eqref{pmssbianchi}, \eqref{pmsseoms} are symmetric under rotations mixing the field strength tensor with its dual,
\be \delta {\cal K}=\ast\, {\cal K}.\ee
The Hodge star operation acts to implement duality, interchanging the roles of the Bianchi identities and the field equations.  The dual gauge field is also a totally symmetric $s$-index tensor, non-locally related to the original gauge field. This is the same phenomenon as electric-magnetic duality in electromagnetism, and holds for all values of $s$ and $t$. 

\section{Depth $t=0$}
\label{sec:maximaldepth}
We begin by considering the simplest case, that of depth $t=0$ partially massless fields (often referred to as ``maximal depth" in other references). These fields possess every helicity component except their scalar polarization. They are  the most direct generalization of electromagnetism and the partially massless spin-2 theory. Here we construct the gauge-invariant curvature tensor for spin-$s> 1$, $t=0$ and show how it reproduces the equations of motion.\footnote{The $s=1$ case is the well-known Maxwell case.  Its analysis is straightforward but doesn't quite fit the general pattern we consider in this section, so we omit it and consider $s> 1$ in the following.}
These tensors are also constructed in~\cite{Gover:2014vxa,Cherney:2015jxp}, and are the higher-spin generalizations of the Maxwell field strength tensor and the PM spin-2 curvature of~\cite{Deser:2006zx}.

The curvature starts out as the $(s+1)$-index tensor,
\be 
{\cal K}_{\mu\nu | \alpha_1\cdots\alpha_{s-1}} \in~\begin{ytableau}
 ~\\~
\end{ytableau}
~~\medotimes~~\begin{array}{|c c c c c|}\hline
&s-1\!\!\!\!\!\!&\!\!\!&&\\
\hline
\end{array}~ , \label{F0rep}
\ee
which is explicitly antisymmetric in its first two indices, explicitly symmetric in its last $s-1$ indices, and has no other symmetry or trace conditions imposed.  We want to show show that the equations of motion \eqref{spinsdepthteom} for a depth $t=0$ PM field of spin-$s$ are equivalent to the equations \eqref{pmssbianchi}, \eqref{pmsseoms} for the tensor ${\cal K}$.

\subsection{Bianchi identities}
We first consider the Bianchi identities \eqref{pmssbianchi}. When we decompose the tensor ${\cal K}_{\mu\nu | \alpha_1\cdots\alpha_{s-1}}$ into irreducible ${\rm GL}(D)$ representations we find the components
\be
\ytableausetup{mathmode}
\ytableausetup{centertableaux,boxsize=1.4em}
\begin{ytableau}
 ~\\~
\end{ytableau}
~~\medotimes~~
\begin{array}{|c c c c c|}\hline
&~&s&~&\\
\hline
\end{array}
~=~
\begin{array}{|c c c c c|}\hline
&\!\!\!\!\!\!\!&s-1&\!\!\!\!\!\!&\\
\hline
~~~\!\!~\vline\!\!\!\\
\cline{1-1}
~~~\!\!~\vline\!\!\!\\
\cline{1-1}
\end{array}
~~\medoplus~~
\begin{array}{|c c c c c|}\hline
&&s&&\\
\hline
~~~\!\!~\vline\!\!\!\\
\cline{1-1}
\end{array}~. \label{maxdepdecompn}
\ee
The algebraic Bianchi identity is the equation ${\rm tr}\ast {\cal K}=0$, which in components reads
\be
\left( \ast {\cal K}\right)_{\mu_1\cdots\mu_{D-3}\rho|~~~\alpha_2\cdots\alpha_{s-1}}^{~~~~~~~~~~~~\rho} \propto \epsilon_{\mu_1\cdots\mu_{D-2}}^{~~~~~~~~~~~\rho\sigma\beta}{\cal K}_{\rho\sigma|\beta\alpha_2\cdots\alpha_{s-1}} = 0.
\ee
Stripping off the epsilon symbol, we find that ${\cal K}$ vanishes if we try to antisymmetrize it over three indices
\be
{\cal K}_{[\mu\nu|\alpha_1]\alpha_2\cdots\alpha_{s-1}} = 0~.
\ee
This means that the component of ${\cal K}$ with the symmetry type \scalebox{.55}{$\begin{array}{|c c c c c|}\hline
&\!\!\!\!\!\!\!&s-1&\!\!\!\!\!\!&\\
\hline
~~~\!\!~\vline\!\!\!\\
\cline{1-1}
~~~\!\!~\vline\!\!\!\\
\cline{1-1}
\end{array}$} vanishes, and thus ${\cal K}$ has the on-shell symmetry
\be
{\cal K}_{\mu\nu,\alpha_1\cdots\alpha_{s-1}} \in~
\begin{array}{|c c c c c|}\hline
&&s&&\\
\hline
~~~\!\!~\vline\!\!\!\\
\cline{1-1}
\end{array} \ \ \ .
\label{Ftenssymmtype}
\ee

Next, we consider the differential Bianchi identity $\rd {\cal K} = 0$, which in components reads
\be
(\rd {\cal K})_{\rho\mu\nu,\alpha_1\cdots\alpha_{s-1}}  \propto \nabla_{[\rho}{\cal K}_{\mu\nu],\alpha_1\cdots\alpha_{s-1}} = 0~.
\label{bianchit0}
\ee
Taking inspiration from electromagnetism and the spin-2 case, we want the operator $\rd$ to be part of a differential complex with zero cohomology. The needed complex is the $t=0$ case of the complex \eqref{depthtspinscomplex}
\be
\label{spinst0complex}
\bullet
\xrightarrow{\overset{\rd^{(s,0)}_1}{\hspace*{.75cm}}}~
\begin{array}{|c c c c c|}\hline
&~&s&~&\\
\hline
\end{array}
\xrightarrow{\overset{\rd^{(s,0)}_2}{\hspace*{.75cm}}}~
\begin{array}{|c c c c c|}\hline
&&s&&\\
\hline
~~~\!\!~\vline\!\!\!\\
\cline{1-1}
\end{array}
\xrightarrow{\overset{\rd^{(s,0)}_3}{\hspace*{.75cm}}}~
\begin{array}{|c c c c c|}\hline
&&s&&\\
\hline
~~~\!\!~\vline\!\!\!\\
\cline{1-1}
~~~\!\!~\vline\!\!\!\\
\cline{1-1}
\end{array}
\xrightarrow{\hspace*{.75cm}}~
\cdots
\ee
where the $\rd^{(s,0)}$ operators maps between the various tensors as
\begin{align}
(\rd^{(s,0)}_1\xi)_{\mu_1\cdots\mu_s}&=\nabla_{(\mu_1}\dots \nabla_{\mu_s)}\xi+\cdots \label{t0gauge} \\\nonumber
&=
\left\{\begin{array}{l}
P_{s}\left(\prod_{n=1}^{\frac{s}{2}}\left[\nabla_{\mu_n}\nabla_{\mu_{n+\frac{s}{2}}}+(2n-1)^2H^2g_{\mu_n\mu_{n+\frac{s}{2}}}\right]\right)\xi~~~~~~~~~{\rm for}~s~{\rm even}\vspace{.15cm}\\
P_{s}\left(\prod_{n=1}^{\frac{s-1}{2}}\left[\nabla_{\mu_n}\nabla_{\mu_{n+\frac{s-1}{2}}}+(2n)^2H^2g_{\mu_n\mu_{n+\frac{s-1}{2}}}\right)\right]\nabla_{\mu_{s}}\xi~~~{\rm for}~s~{\rm odd}
\end{array}\right.\,,\\
(\rd ^{(s,0)}_2\ell)_{\mu\nu,\alpha_1\cdots\alpha_{s-1}} &= 2\nabla_{[\mu}\ell_{\nu]\alpha_1\cdots\alpha_{s-1}} , \label{pmspin2reft1}\\
(\rd ^{(s,0)}_3 {\cal K})_{\mu\nu\rho,\alpha_1\cdots\alpha_{s-1}} &= 3\nabla_{[\rho}{\cal K}_{\mu\nu],\alpha_1\cdots\alpha_{s-1}}\, , \\\nonumber
&~\!~\vdots
\end{align}
Here $\xi$ is a scalar function -- which will be the gauge parameter -- and $\ell_{\alpha_1\cdots\alpha_s}$ is a totally symmetric tensor which will be the fundamental spin-$s$ PM field. The first line,~\eqref{t0gauge}, is the action of the scalar gauge symmetry on $\ell$.
The derivative operator so defined is nilpotent, 
\be \rd^{(s,0)}_{i+1}\circ\rd^{(s,0)}_{i}=0,\ \ \ \ \ \ \ i=1,2,\cdots\, .\ee 
As mentioned, we assume the sequence is exact, in which case the differential Bianchi identity of \eqref{pmssbianchi}, which becomes $\rd^{(s,0)}_{3}{\cal K}=0$ in light of \eqref{Ftenssymmtype}, implies that we may write ${\cal K}$ as the antisymmetric derivative of a totally symmetric tensor $\ell$ as in \eqref{pmspin2reft1},
\be
{\cal K}_{\mu\nu, \alpha_1\cdots\alpha_{s-1}} = 2\nabla_{[\mu}\ell_{\nu]\alpha_1\cdots\alpha_{s-1}}~,
\label{ftensasl}
\ee
which is invariant under a gauge transformation of the form \eqref{t0gauge},
\be \delta \ell_{\mu_1\cdots\mu_s}= (\rd^{(s,0)}_1\xi)_{\mu_1\cdots\mu_s}=\nabla_{(\mu_1}\dots \nabla_{\mu_s)}\xi+\cdots \label{t0gaugesymm3}
\ee
with a scalar gauge parameter, $\xi$.
\subsection{Field equations}
Now that we have the expression \eqref{ftensasl} for the curvature ${\cal K}$ in terms of a symmetric tensor, $\ell$ (which will become the PM field), we want to show that the field equations~\eqref{pmsseoms} reproduce the equations of motion \eqref{spinsdepthteom} for a depth $t=0$ field.
We first consider the algebraic field equation ${\rm tr}\, {\cal K} =0$, which in components reads
\be
 {\cal K}_{\mu\rho,~\!~\alpha_2\cdots\alpha_{s-1}}^{~~\!~~\rho}  = 0.
\ee
Upon using \eqref{ftensasl} this becomes
\be
\nabla_{\mu}\ell_{\alpha_2\cdots\alpha_{s-1}\rho}^{~~~~~~~\!\!~~~~~\rho} - \nabla^\rho\ell_{\rho\mu\alpha_2\cdots\alpha_{s-1}} = 0.
\label{antisymmetricfieldeq}
\ee
This equation has two irreducible components, a fully symmetric part and a mixed symmetry part.  Projecting onto the mixed symmetry part by antisymmetrizing over $\mu$ and $\alpha_2$ we obtain
\be
\nabla_{[\mu}\ell_{\alpha_2]\cdots\alpha_{s-1}\rho}^{~~~~~~~\!\!\!~~~~~~\rho} = 0.
\label{t0trlgauge}
\ee
Comparing this equation to \eqref{pmspin2reft1} tells us that the $\rd^{(s-2,0)}_2$ operator, a member of the complex~\eqref{spinst0complex} with $s\mapsto s-2$, annihilates the trace of $\ell$: $\rd^{(s-2,0)}_2\,{\rm tr}\,\ell=0$.  This, under the assumption of trivial cohomology, implies that ${\rm tr}\,\ell$ is pure gauge, so we can write it as $\rd^{(s-2,0)}_1$ of some scalar function $\chi$,
\bea
\ell_{\alpha_1\cdots\alpha_{s-2}\rho}^{~~~~~~~~~~~\rho}&=&\rd^{(s-2,0)}_1\chi=\nabla_{(\alpha_1}\dots \nabla_{\alpha_{s-2})}\chi+\cdots 
 \label{rfncofh}
\eea

The relation \eqref{rfncofh} is important because it will allow us to set ${\rm tr}\,\ell= 0$ via a gauge choice.  Generically, a scalar gauge freedom would not be expected to be enough to set the trace, a symmetric rank $s-2$ tensor, to zero.\footnote{The exception is the spin-2 case, in which the trace can be gauged directly to zero.}   To set the trace to zero, we would have to find a gauge parameter $\xi$ that solves
\be\label{trace0t0rela}
 \ell_{\alpha_1\cdots\alpha_{s-2}\rho}^{~~~~~~~~~~~\rho} +\delta  \ell_{\alpha_1\cdots\alpha_{s-2}\rho}^{~~~~~~~~~~~\rho} =0.
\ee
However, explicitly evaluating $ \delta\ell_{\alpha_1\cdots\alpha_{s-2}\rho}^{~~~~~~~~~~~\rho} $ given the gauge transformation \eqref{t0gaugesymm3}, we find that the trace of $\ell$ transforms as $\rd^{(s-2,0)}_1$ of a second-order scalar operator $\square+\cdots$ acting on $\xi$, 
\be\label{dlisdboxx}
\delta\ell_{\alpha_1\cdots\alpha_{s-2}\rho}^{~~~~~~~~~~~\rho}=
\left(\rd^{(s-2,0)}_1\left[\square  + H^2(s-1)(D+s-2)\right]\xi\right)_{\alpha_1\cdots\alpha_{s-2}}.
\ee
We now see, using \eqref{rfncofh} and \eqref{dlisdboxx} in \eqref{trace0t0rela}, that the traceless gauge can be reached if we have
\be \rd^{(s-2,0)}_1\left[\chi+\left(\square  + H^2(s-1)(D+s-2)\right)\xi\right]=0.\ee
Invoking the trivial cohomology of \eqref{spinst0complex}, this is satisfied if and only if the term in brackets, $\chi+\left(\square  + H^2(s-1)(D+s-2)\right)\xi,$ vanishes.  
Thus, we can make $\ell$ traceless by taking $\xi$ to satisfy this equation.  This leaves a residual gauge symmetry satisfying the homogeneous equation $\left[\square  + H^2(s-1)(D+s-2)\right]\xi=0$, which is the $t=0$ case of \eqref{genresga}.  

Next we consider the differential equation of motion $\rd\ast \,{\cal K} = 0$, which upon taking another Hodge star and writing in components becomes
\be
\nabla^\nu {\cal K}_{\mu\nu,\alpha_1\cdots\alpha_{s-1}} = 0\,.
\ee
Using~\eqref{ftensasl}, this can be written as
\be
\nabla^\nu {\cal K}_{\mu\nu,\alpha_1\cdots\alpha_{s-1}} \propto \left[\square  - H^2(D+(s-2))\right]\ell_{\mu\alpha_1\cdots\alpha_{s-1}}+(s-1)! H^2 g_{\mu(\alpha_1}\ell_{\alpha_2\cdots\alpha_{s-1})\rho}^{~~~~~~~~~~~~\rho}-\nabla_\mu\nabla_\rho \ell^{\rho}_{~~\alpha_1\cdots\alpha_{s-1}}
 = 0.
\label{sswaveeq}
\ee
Now, in the traceless gauge where $\ell_{\alpha_2\cdots\alpha_{s-1}\rho}^{~~~~~~~~~~~\rho} = 0$, the symmetric part of ~\eqref{antisymmetricfieldeq} tells us that $\ell$ is transverse.  The traceless and transverse conditions along with \eqref{sswaveeq} yield the following system of equations,
\be
\Big(\square  - H^2(D+s-2)\Big)\ell_{\mu\alpha_1\cdots\alpha_{s-1}}=0\,,~~~~~~~~~~\ell_{\alpha_2\cdots\alpha_{s-1}\rho}^{~~~~~~~~~~~\rho} =0\,,~~~~~~~~~~\nabla^\rho\ell_{\rho\mu\alpha_2\cdots\alpha_{s-1}} = 0\,,
\ee
which match \eqref{spinsdepthteom} for a depth $t=0$ PM spin-$s$, along with the residual gauge invariance that preserves the gauge ${\rm tr}\,\ell = 0$, which restricts $\xi$ to satisfy
\be
\Big(\square  + H^2(s-1)(D+s-2)\Big)\xi = 0, \label{residgt0}
\ee
matching \eqref{genresga} in the case $t=0$.

Finally, note that there is a second trace of the curvature tensor we could have taken, ${\cal K}_{\mu\nu,\rho\ \ \alpha_3\cdots\alpha_{s-1}}^{\ \ \ \ \ \rho}$.  This trace is equal to \eqref{t0trlgauge}, so it is automatically constrained to be zero by the original trace requirement ${\rm tr} \,{\cal K} =0$.  The curvature tensor is thus completely traceless on-shell.

\subsection{$d+1$ decomposition\label{t0duality}}

Another way to understand the presence of duality is by looking at how the field strength breaks up under a $D\rightarrow d+1$ decomposition into space and time components.  This is analogous to breaking up the Maxwell field strength into electric and magnetic fields.  On shell, the field strength tensor has the symmetry type \eqref{Ftenssymmtype} and is completely traceless.  Upon reducing $D\rightarrow d+1$, it breaks up into $d$-dimensional fully traceless tensors as (this decomposition follows from the branching rules for the orthogonal groups)
\bea \begin{array}{|c c c c c|}\hline
&&s&&\\
\hline
~~~\!\!~\vline\!\!\!\\
\cline{1-1}
\end{array} ^{\ T}
\xrightarrow[{\scriptscriptstyle D\rightarrow d+1}]{}\hspace{-.5cm}
&& \begin{array}{|c c c c c|}\hline
&&s&&\\
\hline
~~~\!\!~\vline\!\!\!\\
\cline{1-1}
\end{array} ^{\ T}
~~\medoplus~~
\begin{array}{|c c  c |}\hline
&\!\!s-1&\\
\hline
~~~\!\!~\vline\!\!\!\\
\cline{1-1}
\end{array} ^{\ T}
~~\medoplus~\cdots~\medoplus 
~~\begin{ytableau}
 ~\\~
\end{ytableau} \nn\\ 
&& \medoplus~~  
\begin{array}{|c c c c c|}\hline
&~&s&~&\\
\hline
\end{array} ^{\ T}
~~\medoplus
~~ \begin{array}{|c c c|}\hline
&s-1&\\
\hline
\end{array} ^{\ T}
~~\medoplus~\cdots~\medoplus
~~\begin{ytableau}
~
\end{ytableau}~,
\label{Ddbreakupt0}
\eea
where the superscript $T$ indicates that the tableaux are fully traceless.
The spatial tensors in the top line of the right hand side of \eqref{Ddbreakupt0} are analogous to magnetic fields, and those in the bottom line are analogous to electric fields.  

In $D=4$, so that $d=3$, we can dualize the magnetic fields by contracting the anti-symmetric index pair with the spatial epsilon symbol, $\epsilon_{ijk}$, after which they become ordinary traceless symmetric tensors,
\bea 
\begin{array}{|c c c c c|}\hline
&&s&&\\
\hline
~~~\!\!~\vline\!\!\!\\
\cline{1-1}
\end{array} ^{\ T}
\xrightarrow[{\scriptscriptstyle 4\rightarrow 3}]{}\hspace{-.5cm}
&& 
\begin{array}{|c c c c c|}\hline
&~&s&~&\\
\hline
\end{array} ^{\ T}
~~\medoplus~~
\begin{array}{|c c c|}\hline
&s-1&\\
\hline
\end{array} ^{\ T}
~~\medoplus~\cdots~\medoplus~~
\begin{ytableau}
~
\end{ytableau}
\\ 
&& \medoplus~~
\begin{array}{|c c c c c|}\hline
&~&s&~&\\
\hline
\end{array} ^{\ T}
~~\medoplus~~
\begin{array}{|c c c|}\hline
&s-1&\\
\hline
\end{array} ^{\ T}
~~\medoplus~\cdots~\medoplus~~
\begin{ytableau}
~
\end{ytableau} \, , \ \ \  D=4.\nn \label{Ddbreakupt02}
\eea
The electric and magnetic fields now carry the same representation, and the action of duality is to rotate them into each other.  One can think of the magnetic fields as carrying the physical helicity components: $s,s-1,\cdots,1$ of the depth $t=0$ PM field, and the electric fields as carrying their canonical momenta.

\subsection{Example: $s=3$, $t=0$}
The previous discussion was somewhat abstract, so here we work things out explicitly in an example. The spin-2 version was worked out in~\cite{Hinterbichler:2014xga}, so we will do here the next simplest example, spin-3. (For comparison, we have worked out the standard off-shell Lagrangian approach to PM spin-3 in Appendix~\ref{spin3workedout}.)

The starting point is to consider a $4$-index tensor, which is antisymmetric in the first two indices and symmetric in the last two,
\be
{\cal K}_{\mu\nu|\alpha_1\alpha_2} \in~\begin{ytableau}
 ~\\~
\end{ytableau}
~~\medotimes~~
\begin{ytableau}
 ~&~
\end{ytableau}~\,.
\ee
This contains the following components,
\be
\ytableausetup{centertableaux}
\begin{ytableau}
 ~\\~
\end{ytableau}
~~\medotimes~~
\begin{ytableau}
 ~&~
\end{ytableau}
~=~
\begin{ytableau}
 ~&~\\~\\~
\end{ytableau}
~~\medoplus~~
\begin{ytableau}
 ~& ~ &~ \\
~
\end{ytableau}~\,.\label{30exdecompt}
\ee
\paragraph{Bianchi identities:}
Following the general procedure outlined above, we first consider the algebraic Bianchi identity, ${\rm tr}\ast {\cal K}=0$, which implies that the curvature tensor vanishes if we try to antisymmetrize over three indices,
\be
\left({\rm tr}\ast {\cal K}\right)_{\mu_1\cdots\mu_{D-3}\rho|~~\alpha_2}^{~~~~~~\,~~~~~~\rho} \propto \epsilon_{\mu_1\cdots\mu_{D-2}}^{~~~~~~~~~~\rho\sigma\beta}{\cal K}_{\rho\sigma|\beta\alpha_2} = 0 \implies {\cal K}_{[\mu_1\mu_2|\alpha_1]\alpha_2} = 0.
\ee
This means that the part of ${\cal K}$ with the symmetry of the three row Young tableau on the right hand side of \eqref{30exdecompt} vanishes,
and so ${\cal K}$ has the symmetry type ${\cal K}_{\mu\nu,\alpha_1\alpha_2} \in~$\scalebox{.55}{$\begin{ytableau}
 ~& ~ &~ \\
~
\end{ytableau}$}\,.

Next, we consider the differential Bianchi identity $\rd {\cal K}=0$, which in components reads
\be
\nabla_{[\rho}{\cal K}_{\mu\nu],\alpha_1\alpha_2} = 0\,.
\label{s3t0bianchi2}
\ee
This mapping fits into the $s=3$ case of the complex \eqref{spinst0complex}
\be
\label{s3t0complex}
\bullet
\xrightarrow{\overset{\rd_1^{(3,0)}}{\hspace*{.75cm}}}~
\begin{ytableau}
 ~& ~ &~ 
\end{ytableau}
\xrightarrow{\overset{\rd_2^{(3,0)}}{\hspace*{.75cm}}}~
\begin{ytableau}
 ~& ~ &~ \\
~
\end{ytableau}
\xrightarrow{\overset{\rd_3^{(3,0)}}{\hspace*{.75cm}}}~
\begin{ytableau}
 ~& ~ &~ \\
~\\
~
\end{ytableau}
\xrightarrow{\hspace*{.75cm}}~
\cdots,
\ee
where the $\rd$ operator acts as 
\begin{align}
(\rd^{(3,0)}_1\xi)_{\mu_1\mu_2\mu_3} &= \left(\nabla_{(\mu_1}\nabla_{\mu_2}\nabla_{\mu_3)}+4H^2 g_{(\mu_1\mu_2}\nabla_{\mu_3)}\right)\xi \, ,
\label{s3t0oh2terms}\\
(\rd^{(3,0)}_2 \ell)_{\mu_1\mu_2,\alpha_1\alpha_2} &= 2\nabla_{[\mu_1} \ell_{\mu_2],\alpha_1\alpha_2}\, , \label{s3t0oh2terms2} \\
(\rd^{(3,0)}_3 {\cal K})_{\mu_1\mu_2\mu_3,\alpha_1\alpha_2} &=3\nabla_{[\mu_1}{\cal K}_{\mu_2\mu_3],\alpha_1\alpha_2}\, ,\\\nonumber
&~\!~\vdots
\end{align}
One can check straightforwardly that with these definitions, including the ${\cal O}(H^2)$ terms in \eqref{s3t0oh2terms}, the operators satisfy 
\be \rd^{(3,0)}_{i+1}\circ\rd^{(3,0)}_{i}=0,\ \ \  i=1,2,\cdots.  
\ee
As above, we  proceed under the assumption that the sequence~\eqref{s3t0complex} is exact. Then,~\eqref{s3t0bianchi2} becomes $\rd^{(3,0)}_3 {\cal K}=0$ which implies that we can write ${\cal K}$ as an exact form as in \eqref{s3t0oh2terms2} with some symmetric rank-3 tensor $\ell$,\footnote{Note that the spin-3 field that appears in~\eqref{traceshiftedcurvtens} is not the same PM field that appears in the action~\eqref{spin3curvedspacelag}, but is rather a trace-shifted version of it.  The field here may be written in terms of that field as
\begin{equation*}
\ell^{\rm }_{\mu\nu\rho}  = b_{\mu\nu\rho} + \frac{3}{2(D-1)} g_{(\mu\nu}b_{\rho)\alpha}^{~~~\alpha}.
\end{equation*}
Ultimately, the trace is constrained to vanish on-shell, so both fields satisfy the same equations of motion at the end of the day.
}
\be
{\cal K}_{\mu_1\mu_2,\alpha_1\alpha_2} = 2\nabla_{[\mu_1} \ell_{\mu_2]\alpha_1\alpha_2}~.
\label{traceshiftedcurvtens}
\ee
The identity $\rd^{(3,0)}_{2}\circ\rd^{(3,0)}_{1}=0$ then tells us that this curvature tensor is invariant under the gauge symmetry
\be
\label{s3t0gaugesymm}
\delta \ell_{\mu_1\mu_2\mu_3} =  \left(\nabla_{(\mu_1}\nabla_{\mu_2}\nabla_{\mu_3)}+4H^2 g_{(\mu_1\mu_2}\nabla_{\mu_3)}\right)\xi ~,
\ee
for some scalar gauge parameter $\xi$.

\paragraph{Field equations:}
Next we consider the field equations \eqref{pmsseoms}.  The first, algebraic, equation is ${\rm tr}\,{\cal K}=0$, which in components reads
\be
{\cal K}_{\mu\rho~~\alpha}^{~~~\rho}  \propto \nabla_{\mu} \ell_{\alpha\rho}^{~~~\rho} - \nabla^{\rho}\ell_{\rho\mu\alpha} = 0,
\label{traceeqt0}
\ee
while the second, differential, equation $\rd\ast {\cal K}=0$ becomes, upon using \eqref{traceshiftedcurvtens},
\be
\nabla^\nu{\cal K}_{\nu\mu,\alpha_1\alpha_2}=0\implies (\square  - H^2(D+1)) \ell_{\mu\alpha_1\alpha_2}-\nabla_\mu\nabla_\nu \ell^{\nu}_{~~\alpha_1\alpha_2}+2H^2 g_{\mu(\alpha_1} \ell_{\alpha_2)\nu}^{~~\,~~\nu}
 = 0~.
\label{waveeqt0}
\ee

Taking the antisymmetric part of~\eqref{traceeqt0}, we find 
\be
\nabla_{[\mu}\ell_{\alpha]\nu}^{~~~\nu} = 0~,
\ee
which, using the complex \eqref{spinst0complex} in the case $s=1$ (which is nothing but the ordinary de Rham complex), tells us that $\ell_{\mu\nu}^{~~\nu}$ can be written as the gradient of a scalar $\chi$:
\be 
\ell_{\mu\nu}^{~~~\nu} = \nabla_\mu \chi~. \label{trbisscalare}
\ee
This is enough to see that we can choose a gauge where $\ell_{\mu\nu}^{~~\nu} = 0$; to fix this gauge we see from the form of the gauge symmetry \eqref{s3t0gaugesymm} that we need to find a gauge parameter $\xi$ such that
$ \ell_{\mu\nu}^{~~\nu}+\nabla_\mu( \square\xi+2H^2(D+1)\xi)=0,$
which upon using \eqref{trbisscalare} becomes
$ \nabla_\mu\left(\square\xi+2H^2(D+1)\xi+\chi\right)=0.$
Again using the trivial cohomology of our complex \eqref{spinst0complex} in the case $s=1$, the only solution to this is $\square\xi+2H^2(D+1)\xi+\chi=0$, which is solved by a particular solution for $\xi$,  leaving a homogeneous solution which becomes a residual gauge invariance satisfying
$ [\square+2H^2(D+1)]\xi=0.$

In this traceless gauge, \eqref{traceeqt0} tells us that $\ell$ is also transverse, and then the equations of motion \eqref{waveeqt0} become
\be \label{onshellexamt0}
\Big(\square-(D+1)H^2\Big) \ell_{\mu_1\mu_2\mu_3} = 0\,,~~~~~~~\nabla^\nu \ell_{\nu\mu_1\mu_2} = 0\,,~~~~~~~\ell_{\mu\nu}^{~~~\nu} = 0\,,
\ee
which has the residual gauge freedom
\be
\label{s3t0gaugesymmex}
\delta \ell_{\mu_1\mu_2\mu_3} =  \left(\nabla_{(\mu_1}\nabla_{\mu_2}\nabla_{\mu_3)}+4H^2 g_{(\mu_1\mu_2}\nabla_{\mu_3)}\right)\xi\,,~{\rm where}~~ \Big(\square+2(D+1)H^2\Big) \xi = 0.
\ee
These are precisely the same equations as~\eqref{s3maximaldeptheoms}, \eqref{PMspin3symmon}, which describe an on-shell depth $t=0$ PM spin-3.

\section{Depth $t=1$}
\label{sec:deptht1}
The next simplest case to consider is depth $t=1$.  We construct the gauge-invariant curvature tensor for spin-$s> 2$, $t=1$ and show how it reproduces the equations of motion.\footnote{The construction for the massless flat space spin-2 ($t=1$) case is nicely reviewed in~\cite{Bekaert:2002dt}, and the curved space generalization in \cite{Julia:2005ze,Leigh:2007wf}.  It proceeds slightly differently from the general case we consider in this section, so we omit it from the analysis and consider $s> 2$ in the following.}  The depth $t=1$ case will have a curvature tensor construction similar to that of a massless graviton, just as the $t=0$ case behaved as a suitably generalized massless photon.

For $t=1$, the fundamental object is an $(s+2)$-index tensor which has the symmetry type
\be {\cal K}_{\mu_1\nu_1|\mu_2\nu_2,\alpha_1\cdots\alpha_{s-2}} \in~\ytableausetup{centertableaux, boxsize=1.4em}
\begin{ytableau}
 ~\\~
\end{ytableau}
~~\medotimes~~
\begin{array}{|c c c c c|}\hline
&\!\!\!\!\!\!\!&s-1&\!\!\!\!\!\!&\\
\hline
~~~\!\!~\vline\!\!\!\\
\cline{1-1}
\end{array} \ . \label{Rtrepss}
\ee
It is manifestly antisymmetric in the first pair of indices, and the remaining $s$ indices have the symmetry of a tableau with rows of length $s-1$ and $1$, with no additional symmetries or trace conditions.

\subsection{Bianchi identities}
We begin by considering the restrictions that the Bianchi identities~\eqref{pmssbianchi} place on the tensor \eqref{Rtrepss}.  The tensor representation \eqref{Rtrepss} breaks up into the following ${\rm GL}(D)$ irreducible pieces,
\be
\ytableausetup{centertableaux, boxsize=1.4em}
\begin{ytableau}
 ~\\~
\end{ytableau}
~~\medotimes~~
\begin{array}{|c c c c c|}\hline
&\!\!\!\!\!\!\!&s-1&\!\!\!\!\!\!&\\
\hline
~~~\!\!~\vline\!\!\!\\
\cline{1-1}
\end{array}
~=~
\begin{array}{|c c c c c|}\hline
&\!\!\!\!\!\!\!&s-1&\!\!\!\!\!\!&\\
\hline
~~~\!\!~\vline\!\!\!\\
\cline{1-1}
~~~\!\!~\vline\!\!\!\\
\cline{1-1}
~~~\!\!~\vline\!\!\!\\
\cline{1-1}
\end{array}
~~\medoplus~~
\begin{array}{|c c c c c|}\hline
~&\!\!\!s-1\!\!\!\!\!\!\!&~\!\!\!\!\!\!&&\\
\hline
~~~\!\!~\vline\!\!\!&~~~\!\!~\vline\!\!\!\\
\cline{1-2}
~~~\!\!~\vline\!\!\!\\
\cline{1-1}
\end{array}
~~\medoplus~~
\begin{array}{|c c c c c|}\hline
&&s&&\\
\hline
~~~\!\!~\vline\!\!\!\\
\cline{1-1}
~~~\!\!~\vline\!\!\!\\
\cline{1-1}
\end{array}
~~\medoplus~~
\begin{array}{|c c c c c|}\hline
&s\!\!\!\!\!\!\!&&&\\
\hline
~~~\!\!~\vline\!\!\!&~~~\!\!~\vline\!\!\!\\
\cline{1-2}
\end{array}\,.
\label{t1Ryoungtab}
\ee
The first equation of \eqref{pmssbianchi}, the algebraic Bianchi identity ${\rm tr}\ast {\cal K}=0$, gives
\be
\label{t1bianchi1}
\epsilon_{\mu_1\beta_1\cdots\beta_{D-2}}^{~~~~\,~~~~~~~~\beta\rho\sigma}{\cal K}_{\rho\sigma| \beta\nu_2,\alpha_1\cdots\alpha_{s-2}} = 0 \implies {\cal K}_{[\mu_1\nu_1|\mu_2]\nu_2,\alpha_1\cdots\alpha_{s-2}} =0.
\ee
This equation implies that ${\cal K}$ vanishes if we try to antisymmetrize 3 indices, which means that the components on the right hand side of \eqref{t1Ryoungtab} with 3 or more rows vanish. This restricts the symmetry type of ${\cal K}$ to be
\be
{\cal K}_{\mu_1\nu_1,\mu_2\nu_2,\alpha_1\cdots\alpha_{s-2}}\in~\begin{array}{|c c c c c|}\hline
&s\!\!\!\!\!\!\!&&&\\
\hline
~~~\!\!~\vline\!\!\!&~~~\!\!~\vline\!\!\!\\
\cline{1-2}
\end{array}~.
\ee
From this we will see that the two pairs of antisymmetric indices are actually equivalent, i.e., ${\cal K}$ is symmetric under permuting one pair with the other.\footnote{We can now see that it is unnecessary to introduce a second Hodge star or $\rd$ operator acting on the antisymmetric indices in the \scalebox{.55}{$\begin{array}{|c c c c c|}\hline
&\!\!\!\!\!\!\!&s-1&\!\!\!\!\!\!&\\
\hline
~~~\!\!~\vline\!\!\!\\
\cline{1-1}
\end{array}$} factor in~\eqref{Rtrepss}. The second operator would be equivalent to permuting the indices and then using the first operator.} 

The second equation of \eqref{pmssbianchi}, the differential Bianchi identity $\rd {\cal K}=0$, reads
\be
\label{t1bianchi2}
\nabla_{[\rho}{\cal K}_{\mu_1\nu_1],\mu_2\nu_2,\alpha_1\cdots\alpha_{s-2}}=0~.
\ee
This equation fits into the $t=1$ case of the complex \eqref{depthtspinscomplex},
\be
\ytableausetup{centertableaux, boxsize=1.4em}
\begin{ytableau}
~
\end{ytableau}
\xrightarrow{
\overset{\rd^{(s,1)}_1}{\hspace*{.75cm}}}~
\begin{array}{|c c c c c|}\hline
&~&s&~&\\
\hline
\end{array}
\xrightarrow{\overset{\rd^{(s,1)}_2}{\hspace*{.75cm}}}~
\begin{array}{|c c c c c|}\hline
&s\!\!\!\!\!\!\!&&&\\
\hline
~~~\!\!~\vline\!\!\!&~~~\!\!~\vline\!\!\!\\
\cline{1-2}
\end{array}~
\xrightarrow{\overset{\rd^{(s,1)}_3}{\hspace*{.75cm}}}~
\begin{array}{|c c c c c|}\hline
&s\!\!\!\!\!\!\!&&&\\
\hline
~~~\!\!~\vline\!\!\!&~~~\!\!~\vline\!\!\!\\
\cline{1-2}
~~~\!\!~\vline\!\!\!\\
\cline{1-1}
\end{array}
\xrightarrow{\hspace*{.75cm}}~
\cdots
\label{t1complex}
\ee
where the operators act as
\begin{align}
(\rd^{(s,1)}_1\xi)_{\mu_1\cdots\mu_s} &= \nabla_{(\mu_1}\cdots \nabla_{\mu_{s-1}}\xi_{\mu_s)}+\cdots  \label{t2dothterms}\\ \nonumber
&=
\left\{\begin{array}{l}
P_{s}\left(\prod_{n=1}^{\frac{s-1}{2}}\left[\nabla_{\mu_n}\nabla_{\mu_{n+\frac{s-1}{2}}}+(2n-1)^2H^2g_{\mu_n\mu_{n+\frac{s-1}{2}}}\right]\right)\xi_{\mu_{s}}~~~\,{\rm for}~s~{\rm odd}\vspace{.15cm}\\
P_{s}\left(\prod_{n=1}^{\frac{s-2}{2}}\left[\nabla_{\mu_n}\nabla_{\mu_{n+\frac{s-2}{2}}}+(2n)^2H^2g_{\mu_n\mu_{n+\frac{s-2}{2}}}\right]\right)\nabla_{\mu_{s-1}}\xi_{\mu_{s}}~{\rm for}~s~{\rm even}
\end{array}\right.\, , \\ 
(\rd^{(s,1)}_2 \ell)_{\mu_1\nu_1,\mu_2\nu_2,\nu_3\cdots\nu_{s}} &= P_{s,2}\left(\nabla_{\mu_1}\nabla_{\mu_2}+H^2g_{\mu_1\mu_2}\right)\ell_{\nu_1\cdots\nu_s}\, , \label{t2dothterms2} \\
(\rd ^{(s,1)}_3{\cal K})_{\mu_1\nu_1\rho,\mu_2\nu_2,\nu_3\cdots\nu_{s}} &= 3\nabla_{[\rho} {\cal K}_{\mu_1\nu_1]\mu_2\nu_2\nu_3\cdots\nu_{s}}\, ,\\\nonumber
&~~\vdots
\end{align}
The operators defined in this way satisfy 
\be  \rd^{(s,1)}_{i+1}\circ\rd^{(s,1)}_{i}=0,\ \ \ \ \ \ \ i=1,2,\cdots\,, \ee  
and as before, we will assume that the cohomology of this complex is trivial.  With this, the Bianchi identity~\eqref{t1bianchi2} implies that ${\cal K}$ can be written as $\rd$ of an $s$-index symmetric tensor $\ell_{\mu_1\cdots\mu_s}$,
\be
{\cal K}_{\mu_1\nu_1,\mu_2\nu_2,\nu_3\cdots\nu_{s}} = P_{s,2}\left(\nabla_{\mu_1}\nabla_{\mu_2}+H^2g_{\mu_1\mu_2}\right)\ell_{\nu_1\cdots\nu_s}=\left(\rd^{(s,1)}_2\ell\right)_{\mu_1\nu_1,\mu_2\nu_2,\nu_3\cdots\nu_{s}}\, ,
\label{Rintermsofl}
\ee
and is invariant under the gauge transformation
\be
\delta \ell_{\mu_1\mu_2\cdots\mu_s} =\left(\rd^{(s,1)}_1\xi\right)_{\mu_1\cdots\mu_s} =  \nabla_{(\mu_1}\cdots \nabla_{\mu_{s-1}}\xi_{\mu_s)}+\cdots.  
\label{gaugett1gen}
\ee

\subsection{Field equations}
\label{t1fieldeqs}
Having constrained the form of the tensor ${\cal K}$ in terms of $\ell$ by employing the Bianchi identities~\eqref{pmssbianchi}, we now want to show that the evolution equations for the PM field $\ell$ are reproduced by~\eqref{pmsseoms}. We will see that only the equation ${\rm tr}\, {\cal K} = 0$ is necessary.  This equation is already second order in derivatives of $\ell$.  The other equation, $\rd\ast {\cal K} = 0$, is third order in $\ell$, and will become an identity which vanishes by virtue of the second order equations of motion. 

We therefore first consider the algebraic equation, ${\rm tr}\, {\cal K} = 0$, which in components reads
\be
 {\cal K}^\rho_{~\!~\nu_1,\rho\nu_2,\alpha_1\cdots\alpha_{s-2}} =0 .
\ee
Using \eqref{Rintermsofl}, this can be written as
\begin{align}
\nonumber
{\cal K}^\nu_{~\mu_1, \nu\mu_2,\mu_3\cdots\mu_s} \propto \square\ell_{\mu_1\cdots\mu_s}&+(D-2)H^2\ell_{\mu_1\cdots\mu_s} - \nabla_\nu \nabla_{\mu_2} \ell_{\mu_1~\mu_3\mu_4\cdots\mu_s}^{~~\nu}  -\nabla_{\mu_1}\nabla_\nu \ell_{~\mu_2\mu_3\mu_4\cdots\mu_s}^{\nu}\\
&+\nabla_{\mu_1}\nabla_{\mu_2}\ell_{\mu_3\cdots\mu_s\nu}^{~~~~~~~~~\nu}+H^2g_{\mu_1\mu_2}\ell_{\mu_3\cdots\mu_s\nu}^{~~~~~~~~~\nu} = 0\,.
\label{t1spins}
\end{align}
This equation has components of two different symmetry types: a completely symmetric part, and a part with the symmetry of the
Young diagram\footnote{With the exception of $s=3$, where it becomes a hook.  See section \ref{s3t1exsec}. }
\scalebox{.55}{ $
\begin{array}{|c c c c c|}\hline
&s-2\!\!\!\!\!\!\!\!\!\!&&&\\
\hline
~~~\!\!~\vline\!\!\!&~~~\!\!~\vline\!\!\!\\
\cline{1-2}
\end{array}
$}
(with the indices $\mu_1$, $\mu_2$ in the bottom two boxes). If we project~\eqref{t1spins} onto this latter symmetry structure, we find that
the first line vanishes, and the second line becomes the expression for the curvature tensor of the spin-$(s-2)$, $t=1$ field which is the trace $({\rm tr}\, \ell)_{\mu_3\cdots\mu_s}\equiv \ell^\nu_{\ \nu\mu_3\cdots\mu_s}$,
\be  \rd^{(s-2,1)}_2\, \left({\rm tr}\, \ell \right)=0,\ee
where the $\rd^{(s-2,1)}_2$ is as in \eqref{t2dothterms2} with $s\mapsto s-2$.
Therefore, using the complex~\eqref{t1complex} in the case $s\mapsto s-2$, tells us that ${\rm tr}\,\ell$ is pure gauge, and thus can be written as $\rd$ of some vector $\chi_\mu$, as in \eqref{t2dothterms} with the replacement $s\mapsto s-2$,
\be\ell_{\mu_1\cdots\mu_{s-2} \rho}^{~~~~~~~~~~~\rho}= (\rd^{(s-2,1)}_1\chi)_{\mu_1\cdots\mu_{s-2}}=\nabla_{(\mu_1}\cdots \nabla_{\mu_{s-1}}\chi_{\mu_{s-2})}+\cdots\, .
\label{compare1d}
\ee

We next calculate the change in the trace of $\ell$ under a gauge transformation \eqref{gaugett1gen}, and we find two terms, 
\be
\delta\ell_{\mu_1\cdots\mu_{s-2} \rho}^{~~~~~~~~~~~\rho} = \frac{s-2}{s}\left(\rd^{(s-2,1)}_1(\square+H^2[(s-1)(D+s-2)-1])\xi\right)_{\mu_1\cdots\mu_{s-2}}+\frac{2}{s}\left(\rd^{(s-2,1)}_1\left[ \nabla \nabla\cdot\xi\right]\right)_{\mu_1\cdots\mu_{s-2}},
\ee
The first term looks like a spin $s-2$, $t=1$ gauge transformation where the gauge parameter is a the operator $\square+\cdots$ acting on $\xi_\mu$, while in the second term the gauge parameter is $\nabla_\mu\nabla\cdot\xi$.
Comparing to \eqref{compare1d},
and invoking the trivial cohomology of the complex \eqref{t1complex} in the case $s\mapsto s-2$, we see that we can gauge fix the trace of $\ell$ to zero if and only if we choose $\xi_\mu$ to satisfy
\be
\frac{s-2}{s} (\square+H^2[(s-1)(D+s-2)-1])\xi_\mu+ \frac{2}{s}\nabla_\mu\nabla\cdot\xi + \chi_\mu = 0\,.
\ee
This leaves a residual gauge invariance with $\xi_\mu$ satisfying 
\be
(s-2)(\square+H^2[(s-1)(D+s-2)-1])\xi_\mu+ 2\nabla_\mu\nabla\cdot\xi = 0.
\label{residualt1gen}
\ee

Inserting the tracelessness condition back into~\eqref{t1spins} and then anti-symmetrizing over $\mu_1$ and $\mu_3$ we obtain
\be
\nabla_{[\mu_1}\nabla^\nu\ell_{\mu_3]\mu_2\mu_4\cdots\mu_{s-1}\nu} = 0\,.
\ee
This implies, using the complex~\eqref{spinst0complex} in the case $s\mapsto s-1$, that the divergence of $\ell$ can be written in terms of a scalar $\psi$ as
\be
\nabla_\nu\ell^\nu_{~\mu_1\mu_2\cdots\mu_{s-1}} =\left(\rd^{(s-1,0)}_1\psi\right)_{\mu_1\mu_2\cdots\mu_{s-1}}\,.
\label{dlt1coh}
\ee
Now, under a gauge transformation \eqref{gaugett1gen}, the divergence of $\ell$ transforms as a depth-0 field of spin-$(s-1)$ with gauge parameter $\nabla\cdot\xi$:
\be \delta \left(\nabla_\nu\ell^\nu_{~\mu_1\mu_2\cdots\mu_{s-1}} \right)\propto \left(\rd^{(s-1,0)}_1\nabla\cdot \xi\right)_{\mu_1\mu_2\cdots\mu_{s-1}},
\ee
where we have kept in mind that the gauge transformations are now restricted to satisfy the residual equation \eqref{residualt1gen} and used it to eliminate $\square \xi_\mu$ in favor of $\nabla_\mu\nabla\cdot\xi$.
Comparing with \eqref{dlt1coh}, we see that we can use the residual gauge freedom that remains after gauge fixing ${\rm tr}\,\ell = 0$ to gauge-fix the divergence of $\ell$ to be zero as well, by solving $\nabla\cdot\xi\propto \psi$.  This leaves a residual gauge symmetry where the gauge parameter is divergenceless, $\nabla\cdot\xi=0$, in addition to satisfying \eqref{residualt1gen}.

Using the fact that
$
\nabla_\nu \nabla_{\mu_2} \ell_{\mu_1~\mu_3\mu_4\cdots\mu_s}^{~~\nu} = \nabla_{\mu_2} \nabla_{\nu} \ell_{\mu_1~\mu_3\mu_4\cdots\mu_s}^{~~\nu} +(D+(s-2))H^2 \ell_{\mu_1\cdots\mu_s} + {\rm trace\ terms}
$
as well as the divergenceless gauge condition $\nabla^\nu \ell_{\nu\mu_2\cdots\mu_s}=0$, we obtain $\left(\square-sH^2\right)\ell_{\mu_1\cdots\mu_s} = 0$ from~\eqref{t1spins}
so that the equations of motion become
\be
\left(\square-sH^2\right)\ell_{\mu_1\cdots\mu_s} = 0\,,~~~~~~~~~~\nabla^\nu \ell_{\nu\mu_2\cdots\mu_s}=0\,,~~~~~~~~~~\ell^\nu_{\ \nu\mu_3\cdots\mu_{s}}=0\,. \label{t1finalequations}
\ee
where there is a residual gauge invariance of the form~\eqref{t2dothterms} with the gauge parameter satisfying
\be
(\square+H^2[(s-1)(D+s-2)-1])\xi_\mu=0\,~~~~~~~~~~\nabla_\mu\xi^\mu=0\,.
\ee
These equations agree with~\eqref{spinsdepthteom}, \eqref{genresga} for $t=1$.

There is also the second equation of motion in~\eqref{pmsseoms}, $\rd\ast {\cal K} = 0$, which in components becomes
\be \nabla^{\mu_1}{\cal K}_{\mu_1\nu_1,\mu_2\nu_2,\alpha_1\cdots\alpha_{s-2}}=0\,.\ee
Upon inserting \eqref{Rintermsofl}, this turns out to be identically satisfied given the equations \eqref{t1finalequations}.  Finally, we note that the field strength ${\cal K}$ is totally traceless once the equations of motion are satisfied.

\subsection{$d+1$ decomposition}

As we did with the $t=0$ case in section \ref{t0duality}, we can understand the presence of duality by breaking up the field strength into space and time components, analogous to breaking up the Maxwell field strength into electric and magnetic fields. Upon reducing $D\rightarrow d+1$, the traceless on-shell field strength tensor breaks up as
\bea \begin{array}{|c c c c c|}\hline
&s\!\!\!\!\!\!\!\!\!\!&&&\\
\hline
~~~\!\!~\vline\!\!\!&~~~\!\!~\vline\!\!\!\\
\cline{1-2}
\end{array} ^{\ T}
\xrightarrow[{\scriptscriptstyle D\rightarrow d+1}]{}\hspace{-.5cm}
&&\begin{array}{|c c c c c|}\hline
&s\!\!\!\!\!\!\!\!\!\!&&&\\
\hline
~~~\!\!~\vline\!\!\!&~~~\!\!~\vline\!\!\!\\
\cline{1-2}
\end{array} ^{\ T}
~~\medoplus~~
\begin{array}{|c c c c|}\hline
&\!\!\!s-1\!\!\!\!\!\!\!&~&\\
\hline
~~~\!\!~\vline\!\!\!&~~~\!\!~\vline\!\!\!\\
\cline{1-2}
\end{array} ^{\ T}
~~\medoplus~\cdots~\medoplus~~ 
\begin{ytableau}
 ~&~\\~&~
\end{ytableau} ^{\ T} \nn \\
&&
\medoplus~~
\begin{array}{|c c c c c|}\hline
&&s&&\\
\hline
~~~\!\!~\vline\!\!\!\\
\cline{1-1}
\end{array}^{\ T}~~\medoplus~~
\begin{array}{|c c c c|}\hline
&\!\!\!s-1\!\!\!\!\!\!\!&~&\\
\hline
~~~\!\!~\vline\!\!\!&\\
\cline{1-1}
\end{array} ^{\ T}
~~\medoplus~\cdots~\medoplus~~
\begin{ytableau}
 ~&~ \\~
\end{ytableau} ^{\ T}
\nn\\ 
&& \medoplus~~
\begin{array}{|c c c c c|}\hline
&~&s&~&\\
\hline
\end{array}^{\ T}~~\medoplus~~\!
\begin{array}{|c c c c|}\hline
&s-1\!\!\!\!\!\!&~&\\
\hline
\end{array} ^{\ T}~~
\medoplus~\cdots~\!\medoplus
~~\begin{ytableau}
~&~
\end{ytableau}^{\ T}\,. \label{Ddbreakupt1}
\eea

In general $D$ we now have three kinds of spatial field in \eqref{Ddbreakupt1}, not just electric and magnetic fields.  In $D=4$, however, the representations in the first line of the right hand side of \eqref{Ddbreakupt1} carry no independent components and thus are not present, and the representations in the second line 
 can be dualized by contracting the anti-symmetric index pair with the spatial epsilon symbol $\epsilon_{ijk}$, after which they become ordinary symmetric tensors.  This leaves
 \bea \begin{array}{|c c c c c|}\hline
&&s&&\\
\hline
~~~\!\!~\vline\!\!\!\\
\cline{1-1}
\end{array} ^{\ T}
\xrightarrow[{\scriptscriptstyle 4\rightarrow 3}]{}\hspace{-.5cm}
&& 
\begin{array}{|c c c c c|}\hline
&~&s&~&\\
\hline
\end{array} ^{\ T}~~\medoplus~~
\begin{array}{|c c c c|}\hline
&s-1\!\!\!\!\!\!&~&\\
\hline
\end{array}^{\ T}
~~\medoplus~\cdots~\medoplus~~
\begin{ytableau}
~&~
\end{ytableau}^{\ T}
 \\ 
&& \medoplus~~\begin{array}{|c c c c c|}\hline
&~&s&~&\\
\hline
\end{array} ^{\ T}~~\medoplus~~ 
\begin{array}{|c c c c|}\hline
&s-1\!\!\!\!\!\!&~&\\
\hline
\end{array}^{\ T}
 ~~\medoplus~\cdots~\medoplus~~
 \begin{ytableau}
~&~
\end{ytableau} ^{\ T}\,, \ \  D=4.\nonumber \label{Ddbreakupt12}
\eea
The two lines now carry the same representation, and the action of duality is to rotate them into each other.  One can think of these fields as carrying the physical helicity components and canonical momenta for the helicities $s,s-1,\cdots,2$ of the depth $t=1$ PM field.

\subsection{Example: $s=3$, $t=1$\label{s3t1exsec}}
It may be helpful to see the construction of the previous section worked out in an explicit example. The simplest field which admits a depth $t=1$ partially massless point is a spin-3.  We will work out this case in this section.  We have worked out the off-shell Lagrangian approach in Appendix~\ref{spin3workedout} for comparison.

The starting point is to consider the tensor
\be
{\cal K}_{\mu_1\nu_1|\mu_2\nu_2,\alpha}\in~\begin{ytableau}
 ~\\~
\end{ytableau}
~~\medotimes~~
\begin{ytableau}
 ~&~\\~
\end{ytableau}\,,
\label{s3t1curvature}
\ee
which has the following ${\rm GL}(D)$ decomposition
\be
\label{s3t1youngdecomp}
\ytableausetup{centertableaux, boxsize=1.4em}
\begin{ytableau}
 ~\\~
\end{ytableau}
~~\medotimes~~
\begin{ytableau}
 ~&~\\~
\end{ytableau}
~=~
\begin{ytableau}
 ~&~\\~\\~\\~
\end{ytableau}
~\medoplus~~
\begin{ytableau}
 ~&~&~\\~\\~
\end{ytableau}
~~\medoplus~~
\begin{ytableau}
 ~& ~&~ \\
~&~
\end{ytableau}~.
\ee
\paragraph{Bianchi identities:}
We first consider the constraints that the Bianchi identities place on the form of the tensor~\eqref{s3t1curvature}. The algebraic Bianchi identity ${\rm tr}\ast {\cal K}=0$ implies (after stripping of an epsilon) that ${\cal K}$ vanishes if we try to antisymmetrize over the first three of its indices,
\be {\cal K}_{[\mu_1\nu_1|\mu_2]\nu_2,\alpha} = 0.\ee
Therefore the components of the decomposition~\eqref{s3t1youngdecomp} with more than two rows vanish, so that the tensor ${\cal K}$ has the symmetry type
\be {\cal K}_{\mu_1\nu_1|\mu_2\nu_2,\alpha}\in\,\,
\begin{ytableau}
 ~&~&~\\~&~
\end{ytableau}
\,.
\ee

Next, we consider the differential Bianchi identity $\rd {\cal K}=0$, which in components reads
\be
\nabla_{[\rho}{\cal K}_{\mu_1\nu_1],\mu_2\nu_2,\alpha}=0\,. \label{bianchis3t1}
\ee
This identity fits into the complex
\be
\ytableausetup{centertableaux, boxsize=1.4em}
\begin{ytableau}
~
\end{ytableau}
\xrightarrow{
\overset{\rd^{(3,1)}_1}{\hspace*{.75cm}}}~
\begin{ytableau}
 ~& ~ &
\end{ytableau}
\xrightarrow{\overset{\rd^{(3,1)}_2}{\hspace*{.75cm}}}~
\begin{ytableau}
 ~& ~ &\\
 ~&~
\end{ytableau}
\xrightarrow{\overset{\rd^{(3,1)}_3}{\hspace*{.75cm}}}~
\begin{ytableau}
 ~& ~ &\\
 ~&~\\
 ~
\end{ytableau}
\xrightarrow{\hspace*{.75cm}}~
\cdots
\label{s3t1complex}
\ee
where the differential operators act as
\begin{align}
\label{s3t1gaugetrans}
(\rd^{(3,1)}_1\xi)_{\mu_1\mu_2\mu_3} &= \nabla_{(\mu_1}\nabla_{\mu_2}\xi_{\mu_3)}+H^2 g_{(\mu_1\mu_2}\xi_{\mu_3)}\, ,\\\label{s3t1curv}
(\rd ^{(3,1)}_2\ell)_{\mu_1\nu_1,\mu_2\nu_2,\mu_3} &= P_{3,2}\left(\nabla_{\nu_1}\nabla_{\nu_2}+H^2g_{\nu_1\nu_2}\right)\ell_{\mu_1\mu_2\mu_3}\, ,\\
(\rd ^{(3,1)}_3 {\cal K})_{\mu_1\nu_1\rho,\mu_2\nu_2,\mu_3} &= 3\nabla_{[\rho} {\cal K}_{\nu_1\mu_1],\nu_2\mu_2,\mu_3}\, ,\\\nonumber
&~\!~\vdots
\end{align}
It is straightforward to check that with these definitions we have 
\be \rd^{(3,1)}_{i+1}\circ\rd^{(3,1)}_{i}=0, \ \  i=1,2,\cdots.
\ee
Using this complex and the assumption of trivial cohomology, the Bianchi identity \eqref{bianchis3t1}, which is $\rd ^{(3,1)}_3 {\cal K} = 0$, implies that we can write ${\cal K}$ in terms of a potential as
\be
{\cal K}_{\mu_1\nu_1,\mu_2\nu_2,\mu_3} = ( \rd ^{(3,1)}_2 \ell)_{\mu_1\nu_1,\mu_2\nu_2,\mu_3} = P_{3,2}\left(\nabla_{\nu_1}\nabla_{\nu_2}+H^2g_{\nu_1\nu_2}\right)\ell_{\mu_1\mu_2\mu_3}~, \label{s3t1fintp}
\ee
which, by virtue of $\rd^{(3,1)}_{2}\circ\rd^{(3,1)}_{1}=0$, is gauge invariant under the gauge transformation of the form~\eqref{s3t1gaugetrans},
\be
\delta \ell_{\mu_1\mu_2\mu_3} = \nabla_{(\mu_1}\nabla_{\mu_2}\xi_{\mu_3)}+H^2 g_{(\mu_1\mu_2}\xi_{\mu_3)}\,.
\ee
\paragraph{Equations of motion:}
The trace equation ${\rm tr}\ast {\cal K}=0$ becomes, upon using \eqref{s3t1fintp},
\be
{\cal K}^{\ \ \rho}_{\mu_1\ ,\rho\mu_2,\mu_3} \propto \big(\square  -3 H^2\big)\ell_{\mu_1\mu_2\mu_3}-2\nabla_{(\mu_1}\nabla_{\lvert\rho\rvert} \ell^\rho_{~\mu_2)\mu_3}+\nabla_{\mu_2}\nabla_{\mu_1} \ell_{\mu_3\rho}^{~~~~\rho} +2H^2 g_{\mu_1\mu_2} \ell_{\mu_3\rho}^{~~~~\rho}+H^2 g_{\mu_1\mu_3}\ell_{\mu_2\rho}^{~~~~\rho}= 0\,.
\label{s3t1traceeq}
\ee
This equation has two parts, a totally symmetric part and a part with the symmetry~\scalebox{.55}{$
\begin{ytableau}
 ~&~\\~
\end{ytableau}
$}\,. Projecting~\eqref{s3t1traceeq} onto the hook tableau (which amounts to antisymmetrizing over $\mu_1,\mu_3$), we obtain the equation
\be
2H^2 g_{\mu_2[\mu_1} \ell_{\mu_3]\rho}^{~~~~\rho}-\nabla_{[\mu_1}\nabla_{\lvert\rho\rvert} \ell^\rho_{~\mu_3]\mu_2}+\nabla_{\mu_2}\nabla_{[\mu_1} \ell_{\mu_3]\rho}^{~~~~\rho} = 0\,.
\label{s3t1eomhook}
\ee

Under a gauge transformation, the trace of $\ell$ transforms into
\be
 \ell_{\mu\rho}^{~~~\rho}\mapsto \ell_{\mu\rho}^{~~~\rho} + \frac{1}{3}\left(\square\xi_\mu+(1+2D)H^2\right)\xi_\mu+\frac{2}{3}\nabla_\mu \nabla\cdot\xi\,.
\ee
We may fix the traceless gauge $\ell_{\mu\rho}^{~~~\rho}=0$ by setting the above to zero and solving the resulting differential equation for $\xi$.\footnote{Note that this is somewhat simpler than the general case, where there would be another projection of~\eqref{s3t1traceeq} which would tell us that we have enough gauge freedom to fix ${\rm tr}~\ell = 0$. In this case, no such argument is needed because the trace is a vector, which can directly be fixed to zero with our vector gauge symmetry.}  This leaves a residual gauge symmetry given by the homogeneous solution,
\be  
\left(\square+(1+2D)H^2\right)\xi_\mu+2\nabla_\mu \nabla\cdot\xi=0.\label{residt1s3ex}
\ee
Using this gauge choice in~\eqref{s3t1eomhook} tells us that
\be
\nabla_{[\mu_1}\nabla_{\lvert\rho\rvert} \ell^\rho_{~\mu_3]\mu_2} =0.
\ee
We recognize this as the operator $\rd^{(2,0)}_2$ in the complex~\eqref{spinst0complex}, annihilating the symmetric tensor $\nabla_\rho \ell^\rho_{~~\mu_1\mu_2}$.  Using the trivial cohomology assumption, this implies that the divergence of $\ell$ can be written as $\rd^{(2,0)}_1$ of some scalar, $\chi$,
\be
\nabla_\lambda \ell^\lambda_{~\!~\mu_1{\mu_2}} = (\rd^{(2,0)}_1 \chi)_{\mu_1{\mu_2}} = (\nabla_{\mu_1}\nabla_{\mu_2}+H^2g_{\mu_1{\mu_2}})\chi.\label{divbpureg}
\ee

Next we examine how the divergence of $\ell$ transforms under a gauge transformation,
\be
\delta\nabla_\lambda \ell^\lambda_{~\!~\mu_1{\mu_2}} = \frac{2}{3}\nabla_{(\mu_1}\left(\square+H^2(1+2D)\right)\xi_{\mu_2)}+ \frac{1}{3}\left(\nabla_{\mu_1}\nabla_{\mu_2}-3H^2g_{\mu_1{\mu_2}}\right)\nabla\cdot\xi.
\label{divgauges3}
\ee
We would like to fix the divergence to be zero using only the residual gauge symmetry satisfying \eqref{residt1s3ex}.  
Using \eqref{residt1s3ex} in~\eqref{divgauges3}, we find that the divergence of $\ell$ transforms as
\be
\nabla_\lambda \ell^\lambda_{~\!~\mu_1{\mu_2}}\mapsto\nabla_\lambda \ell^\lambda_{~\!~\mu_1{\mu_2}}   -\left(\nabla_{\mu_1}\nabla_{\mu_2}+H^2 g_{\mu_1{\mu_2}}\right)\nabla\cdot\xi,
\ee
under this residual gauge symmetry. 
Therefore, using \eqref{divbpureg}, and the trivial cohomology assumption of the complex \eqref{spinst0complex}, we see that we can
 also gauge-fix $\nabla_\lambda \ell^\lambda_{~\!~\mu_1{\mu_2}} = 0$ if and only if $\nabla\cdot\xi=\chi$. Once we have fixed this gauge, there is still a residual gauge freedom, where $\xi$ is divergenceless $\nabla\cdot\xi = 0$, and satisfies the equation $\left(\square+H^2(1+2D)\right)\xi_\mu = 0$.

After inserting the two conditions $\nabla_\lambda \ell^\lambda_{~\!~\mu_1{\mu_2}} = 0$ and $\ell_{\mu\rho}^{~\!~~\rho} = 0$ into~\eqref{s3t1traceeq} we obtain the wave equation $ \left(\square  -3 H^2\right)\ell_{\mu_1\mu_2\mu_3} = 0$. Thus the equations of motion following from ${\rm tr}\,{\cal K} = 0$ are
\be
 \left(\square  -3 H^2\right)\ell_{\mu_1\mu_2\mu_3}=0~~~~~~~~~~~~\ell_{\mu\rho}^{~\!~~\rho} = 0~~~~~~~~~~~~\nabla_\lambda \ell^\lambda_{~\!~\mu{\nu}} = 0~, \label{spin3t1eomr1}
\ee
which are invariant under the residual gauge symmetry 
\bea
\delta \ell_{\mu_1\mu_2\mu_3} = \nabla_{(\mu_1}\nabla_{\mu_2}\xi_{\mu_3)}+H^2 g_{(\mu_1\mu_2}\xi_{\mu_3)}\,,~~{\rm where}~~ \left(\square+H^2(1+2D)\right)\xi_\mu = 0\,,\ \ \ \nabla\cdot\xi = 0\,.   \label{spin3t1eomr2}
\eea
These is precisely the $s=3$, $t=1$ case of the partially massless on-shell equations \eqref{spinsdepthteom} and on-shell gauge symmetries \eqref{genresga}.

One can check directly that the remaining equation $\rd\,\ast {\cal K} = 0\implies \nabla^{\mu_1}{\cal K}_{\mu_1\mu_1,\mu_2\mu_2,\mu_3}=0$ does not provide any new information, but is satisfied identically once \eqref{s3t1traceeq} is satisfied (this is true even without fixing a gauge, as can be seen by using the trace of~\eqref{s3t1eomhook} to eliminate $\square\ell_{\mu\rho}^{~\,~\rho}$ in $\nabla\cdot{\cal K}$, after which the result is proportional to~\eqref{s3t1traceeq}).  Finally, it can be checked that the field strength is fully traceless once \eqref{spin3t1eomr1} are satisfied.

\section{Depth $t\geq2$}
\label{sec:higherdepth}
We have seen that for higher spins, the depth $t=0$ case is reminiscent of the structure of a massless spin-1 field in the sense that the field strength tensor ${\cal K}$ is first order in derivatives and the second order equations of motion follow from the one-derivative equation $\rd\ast {\cal K} = 0$. The depth $t=1$ case shares many of the features of a massless spin-2 field, in which the field strength tensor ${\cal K}$ is second order in derivatives and the second order equations of motion follow from the zero-derivative equation ${\rm tr}\, {\cal K}  = 0$. 
We therefore suspect that a depth-$t$ PM theory will behave similarly to a massless spin-$(t+1)$ field.  But in this case, the curvature tensor has $\geq 3$  derivatives and it naively seems impossible for the second order equations to arise from either of the dynamical equations \eqref{pmsseoms}.
They do in fact arise, but the equations of motion will be implemented via gauge fixing, similar to the way in which the equations for a massless spin-$s$ appear in~\cite{Francia:2002aa,Francia:2002pt,Bekaert:2003az,Sagnotti:2003qa} (reviewed in~\cite{Sorokin:2004ie,Bouatta:2004kk}).

Here we sketch how the construction works for depth-$t$ PM theories,\footnote{We restrict to $s<s-1$ because $t=s-1$ is the massless case, for which the details are somewhat different and which is a straightforward curved space generalization of what has been done before in the flat case.} with $t\geq 2$, $t<s-1$.  The starting point is an $(s+t+1)$-index tensor ${\cal K}$ with the symmetry type:
\be
{\cal K}_{\mu_1\nu_1|\mu_2\nu_2,\cdots,\mu_{t+1}\nu_{t+1},\alpha_1\cdots\alpha_{s-t-1}}\in~\ytableausetup{centertableaux,boxsize=1.4em}
\begin{ytableau}
 ~\\~
\end{ytableau}
~~\medotimes~~\begin{array}{|c c c c c|}\hline
&s-1\!\!\!\!\!\!&\!\!\!&&\\
\hline
\!\!\!  &\!\! t~~~  \vline\!\!\!\!\!\!\\
\cline{1-2}
\end{array}~.
\label{spinsdeptht}
\ee
The Hodge dual and exterior $\rd$ operations are again defined with respect to the antisymmetric pair of indices in the first factor on the right hand side of \eqref{spinsdeptht},\footnote{As mentioned in a previous footnote, there are in principle multiple ways that we could dualize the tensor ${\cal K}$, but they will all end up being equivalent.  Choosing to only dualize along the first antisymmetric indices makes contact with the frame-like formulation of~\cite{Skvortsov:2006at}. Similarly, we could introduce $t+1$ differential operators and consider ${\cal K}$ as a multiform, but this is not necessary for our construction.} as described in section \ref{fieldseomsubsec}.
The tensor ${\cal K}$ is the PM version of the curvature tensors considered in~\cite{Francia:2002aa,Francia:2002pt,Bekaert:2003az}, which are based on the higher-spin curvatures constructed in~\cite{deWit:1979pe,Damour:1987vm}.

\subsection{Bianchi identities}
We first explore the consequences of the Bianchi identities for the structure of the tensor ${\cal K}$.  The trace constraint, ${\rm tr}\ast{\cal K}=0$, becomes in components
\be \epsilon_{\mu_1\beta_1\cdots\beta_{D-2}}^{~~~~~~~~~~~~~\beta\rho\sigma}{\cal K}_{\rho\sigma| \beta\nu_2,\cdots,\mu_{t+1}\nu_{t+1},\alpha_1\cdots\alpha_{s-t-1}} = 0 \implies {\cal K}_{[\mu_1\nu_1|\mu_2]\nu_2,\cdots,\mu_{t+1}\nu_{t+1},\alpha_1\cdots\alpha_{s-t-1}} =0.
\label{traceequationgent}
\ee
The representation \eqref{spinsdeptht} decomposes under ${\rm GL}(D)$ as
\be \begin{ytableau}
 ~\\~
\end{ytableau}
~~\medotimes~~\begin{array}{|c c c c c|}\hline
&s-1\!\!\!\!\!\!&\!\!\!&&\\
\hline
\!\!\!  &\!\! t~~~  \vline\!\!\!\!\!\!\\
\cline{1-2}
\end{array}~=~
\begin{array}{|c c c c c|}\hline
~&~s~~~~\!\!\!\!\!\!&\!\!\!&&\\
\hline
\!\!\!  &\!\!\!\!\!~t~~~~ \vline\!\!\!\!\!\\
\cline{1-2}
~~~\!\!~\vline\!\!\!\\
\cline{1-1}
\end{array}
~~\medoplus~~
\begin{array}{|c c c c c|}\hline
\!\!&\!\!\!\!s-1\!\!\!\!&\!\!\!\!&\!\!&\\
\hline
\!\!\!  \!\!\!&\!\!\!\!\!\! t+1~~  \vline\!\!\!\\
\cline{1-2}
~~~\!\!~\vline\!\!\!\\
\cline{1-1}
\end{array} 
~~\medoplus~~ 
\begin{array}{|c c c c c|}\hline
\!\!&\!\!\!\!s-1\!\!\!\!&\!\!\!\!&\!\!&\\
\hline
\!\!\!  \!\!\!&\!\!\!\!\!\! t+1~~  \vline\!\!\!\\
\cline{1-2}
~~~\!\!~\vline\!\!\!\\
\cline{1-1}
~~~\!\!~\vline\!\!\!\\
\cline{1-1}
\end{array} 
 ~~\medoplus~~ 
\begin{array}{|c c c c c|}\hline
&s\!\!\!\!\!\!&\!\!\!&&\\
\hline
\!\!\!  &\!\!\! t+1~~  \vline\!\!\!\\
\cline{1-2}
\end{array} ~,
\label{gentydecomp}
\ee
and \eqref{traceequationgent} implies
that all of the components on the right hand side of \eqref{gentydecomp} which have $3$ or more rows are zero. The only component which survives this projection is the one with the symmetry type\footnote{Note that this tensor has the same symmetry type as the projected curvature tensor of~\cite{Skvortsov:2006at}.}
\be 
{\cal K}_{\mu_1\nu_1,\cdots,\mu_{t+1}\nu_{t+1},\alpha_1\cdots\alpha_{s-t-1}}\in~ 
\begin{array}{|c c c c c|}\hline
&s\!\!\!\!\!\!&\!\!\!&&\\
\hline
\!\!\!  &\!\!\! t+1~~  \vline\!\!\!\\
\cline{1-2}
\end{array}\,.
\label{spinsdepthtcurv}
\ee

 We now turn to the differential Bianchi identity, $\rd\,{\cal K}=0$, which reads
\be
\label{t2bianchi2}
\nabla_{[\rho}{\cal K}_{\mu_1\nu_1],\mu_2\nu_2,\cdots,\mu_{t+1}\nu_{t+1},\alpha_1\cdots\alpha_{s-t-1}}=0\,.
\ee
This differential operator is $\rd^{(s,t)}_3$, part of the complex \eqref{depthtspinscomplex} with the operators defined as in \eqref{gendc1}-\eqref{gendc3}.
Again assuming that the cohomology of this complex is trivial, the fact that ${\cal K}$ is annihilated by $\rd^{(s,t)}_3$ implies that it may be written as $\rd^{(s,t)}_2$ acting on a totally symmetric rank-$s$ potential $\ell$ as
\be
{\cal K}_{\mu_1\cdots\mu_{s},\nu_1\cdots \nu_{t+1}} ={\rd^{(s,t)}_2 \ell}_{\mu_1\cdots\mu_{s},\nu_1\cdots \nu_{t+1}}
\propto P_{s,t+1}\nabla_{\nu_{1}}\cdots\nabla_{\nu_{t+1}}\ell_{\mu_{1}\cdots\mu_{s}}+\cdots \ \,.
\label{kisspinstg}
\ee
By virtue of $\rd^{(s,t)}_2\circ \rd^{(s,t)}_1=0$, this curvature tensor is gauge invariant under a transformation of the form
\be
\delta\ell_{\mu_1\cdots \mu_s}
= \left(\rd^{(s,t)}_1 \xi\right)_{\mu_1\cdots \mu_s}\propto \nabla_{(\mu_{t+1}}\nabla_{\mu_{t+2}}\cdots\nabla_{\mu_{s}}\xi_{\mu_{1}\cdots\mu_{t})}+\cdots 
\ee
for a totally symmetric $t$ index gauge parameter $\xi$. The explicit expressions for the curvatures and gauge transformations, including the sub-leading ${\cal O}(H^2)$ terms are given in section~\ref{diffcomplexsect}.

\subsection{Field equations}
\label{anydepthfieldeqs}

The combined effect of the Bianchi identities is to restrict the tensor~\eqref{spinsdeptht} to be of the symmetry type~\eqref{spinsdepthtcurv} written in terms of a spin-$s$ potential as in \eqref{kisspinstg}. What remains is to show that we can recover the equations of motion from this curvature using the field equations \eqref{pmsseoms}. A priori this is impossible, because the equations of motion are manifestly second order, whereas the curvature tensor \eqref{kisspinstg} has $t+1$ derivatives, which is $\geq 3$ since $t\geq2$. However, we will see that similar to~\cite{Francia:2002aa,Francia:2002pt,Bekaert:2003az}, the equation ${\rm tr}\,{\cal K} = 0$, plus appropriate gauge fixings, does indeed imply the correct partially massless equations of motion.

We begin by considering the algebraic equation of motion ${\rm tr}\, {\cal K}=0$,
\be
{\cal K}^\rho_{~\nu_1,\rho\nu_2,\cdots,\mu_{t+1}\nu_{t+1},\alpha_{1}\cdots\alpha_{s-1-t}}= 0.
\label{traceeq}
\ee
This equation contains three distinct symmetry components
\be
{\rm tr}\,{\cal K} ~\supset~~ \begin{array}{|c c c c c|}\hline
&s\!\!\!\!\!\!&\!\!\!&&\\
\hline
\!\!\!  &\!\!\! t-1~~  \vline\!\!\!\\
\cline{1-2}
\end{array}
~~\medoplus~~
\begin{array}{|c c c c c|}\hline
&s-2\!\!\!\!\!&\!\!\!\!\!\!\!&&\\
\hline
\!\!\!  &\!\!\! t+1~~  \vline\!\!\!\\
\cline{1-2}
\end{array}
~~\medoplus~~
\begin{array}{|c c c c c|}\hline
&s-1\!\!\!\!\!&\!\!\!\!\!\!\!&&\\
\hline
\!\!\!  &\!\! t~~~\vline\!\!\!\!\!\!\!\\
\cline{1-2}
\end{array}\,, \label{trksymt2genc}
\ee
and we will get various equations by projecting onto each of these three components (the second component is absent in the case $t=s-2$). 

We first consider the projection onto the Young diagram
\scalebox{.55}{
$
\begin{array}{|c c c c c|}\hline
&s\!\!\!\!\!\!&\!\!\!&&\\
\hline
\!\!\!  &\!\!\! t-1~~  \vline\!\!\!\\
\cline{1-2}
\end{array}
$
}.
As it turns out, the result can be written as $\rd^{(s,t-2)}_2$ of a totally symmetric rank $s$ tensor ${\cal F}_{\alpha_1\cdots \alpha_s}$, so that we have
\be
\rd^{(s,t-2)}_2{\cal F}=0 ~.
\label{ftensdoperator}
\ee
The tensor ${\cal F}$ is the PM analogue of the Fronsdal tensor, and contains only up to second derivatives of $\ell$.  The additional $t-1$ derivatives needed to obtain the $t+1$ derivative field strength are all present in the operator $\rd^{(s,t-2)}_2$ in \eqref{ftensdoperator}.  The relation \eqref{ftensdoperator} is a generalization of a similar identity noted in the massless spin-3 case \cite{Damour:1987vm}.

The goal is to show that ${\cal F}$ is zero, from which the standard on-shell equations will follow after appropriate gauge-fixings. 
\eqref{ftensdoperator} implies, using the complex \eqref{depthtspinscomplex} with $t\mapsto t-2$, that we can write ${\cal F}$ as $\rd^{(s,t-2)}_1$ of some rank $t-2$ totally symmetric tensor $\Lambda_{\alpha_1\cdots\alpha_{t-2}}$, 
\be {\cal F}=\rd^{(s,t-2)}_1\Lambda.\label{gengtf4}\ee
Next, we note that under a gauge transformation~\eqref{kisspinstg}, ${\cal F}$ transforms as
\be \delta {\cal F}=\rd^{(s,t-2)}_1\,{\rm tr}\, \xi.\ee
Comparing with \eqref{gengtf4}, we see that we can use the trace of the gauge symmetry to gauge fix 
\be {\cal F} = 0\ .\label{Fvanishesgen}\ee
This is the same mechanism by which the Fronsdal equations of motion are recovered in the flat-space massless case in \cite{Francia:2002aa,Francia:2002pt,Bekaert:2003az}. 
This leaves a residual gauge symmetry with a traceless gauge parameter,
\be {\rm tr}\, \xi=0.\ee
 (Note that the tensor ${\cal F}$ is invariant under the gauge symmetry~\eqref{kisspinstg} with a {traceless} gauge parameter, just as in the standard Fronsdal formulation of the massless case.)

We next consider the projection of \eqref{traceeq} onto the Young diagram~\scalebox{.55}{$
\begin{array}{|c c c c c|}\hline
&s-1\!\!\!\!\!&\!\!\!\!\!\!\!&&\\
\hline
\!\!\!  &\!\! t~~~\vline\!\!\!\!\!\!\!\\
\cline{1-2}
\end{array}$
}.
The result can be written as $\rd^{(s-1,t-1)}_2$ acting on a de Donder-like linear combination of the divergence $\left(\nabla\cdot \ell\right)_{\mu_1\cdots\mu_{s-1}} \equiv\nabla^\rho\ell_{\rho\mu_1\cdots\mu_{s-1}}$ and the symmetrized derivative of the trace $\left(\nabla {\rm tr}\ell\right)_{\mu_1\cdots\mu_{s-1}}\equiv \nabla_{(\mu_1}{\rm tr}\,\ell_{\mu_2\cdots\mu_{s-1})}$,
\be
\rd^{(s-1,t-1)}_2\left(\nabla\cdot\ell -  \frac{s-1}{s-1-t}\nabla {\rm tr}\,\ell\right) = 0.
\ee
The fact that this linear combination is annihilated by $\rd^{(s-1,t-1)}_2$ implies that we can write is as $\rd^{(s-1,t-1)}_1$ of some rank $t-1$ fully symmetric tensor $\psi$
\be
\left(\nabla\cdot\ell -  \frac{s-1}{s-1-t}\nabla\,{\rm tr}\,\ell\right) = \rd^{(s-1,t-1)}_1\psi.
\ee
Under a gauge transformation, this same linear combination transforms as
\be
\delta\left(\nabla\cdot\ell -  \frac{s-1}{s-1-t}\nabla\,{\rm tr}\,\ell\right) = \rd^{(s-1,t-1)}_1\nabla\cdot\xi,
\ee
so we see that we can use the divergence of the gauge parameter to fix the de Donder type gauge where 
\be \nabla\cdot\ell -  \frac{s-1}{s-1-t}\nabla {\rm tr}\,\ell = 0.\label{dedondergenst}\ee   This leaves residual gauge transformations where the gauge parameter is transverse $\nabla\cdot\xi = 0$.

Finally, we project the equation of motion~\eqref{traceeq} onto the symmetry structure
\scalebox{.55}{$
\begin{array}{|c c c c c|}\hline
&s-2\!\!\!\!\!&\!\!\!\!\!\!\!&&\\
\hline
\!\!\!  &\!\!\! t+1~~  \vline\!\!\!\\
\cline{1-2}
\end{array}$
},
and we find that the equation becomes
\be
\rd^{(s-2,t)}_2 \, {\rm tr}\, \ell  = 0\,,
\ee
which implies that the trace of $\ell$ can be written as $\rd^{(s-2,t)}_1$ of some rank $t$ fully symmetric tensor $\chi$,
\be {\rm tr}\, \ell  =\rd^{(s-2,t)}_1\chi.\label{trellchigen}\ee
(In the case $t=s-2$ we do not have this equation, and we do not need this cohomology argument for ${\rm tr}\,\ell$, so we define $\chi\equiv {\rm tr}\, \ell$ in this case, see the example in section \ref{s4t2section}.)
Under a gauge transformation, the trace of $\ell$ shifts as $\delta\,{\rm tr}\, \ell = {\rm tr}\,\rd^{(s,t)}_1\xi\,$, which upon using that the gauge parameter is transverse and traceless can be cast as  
\be \delta\,{\rm tr}\, \ell =  \rd^{(s-2,t)}_1\left(\square+ H^2\left[(s-1)(D+s-2)-t\right]\right)\xi.\ee
Comparing with \eqref{trellchigen} and using the trivial cohomology assumption, we can reach the gauge ${\rm tr}\, \ell=0$ if we can solve the equation $\left(\square+ H^2\left[(s-1)(D+s-2)-t\right]\right)\xi\propto \chi$ with a transverse traceless $\xi$, which would leave a residual gauge transformation satisfying the homogeneous equation $\left(\square+ H^2\left[(s-1)(D+s-2)-t\right]\right)\xi=0$.

This will be possible if $\chi$ is itself transverse and traceless, so we now turn to arguing that this is indeed the case.  Start by plugging \eqref{trellchigen} into the trace of the de Donder condition \eqref{dedondergenst}.  The result can be cast as $\rd^{(s-3,t-1)}_1$ acting on a de Donder-like expression for $\chi$, 
\be {\rm tr} \left(\nabla\cdot\ell -  \frac{s-1}{s-1-t}\nabla {\rm tr}\,\ell\right)\propto \rd^{(s-3,t-1)}_1\left(\nabla\cdot\chi +\frac{t-1}{s-1-t}\nabla {\rm tr}\,\chi\right) = 0,\ee
Thus by the trivial cohomology assumption and \eqref{dedondergenst} we have 
\be \nabla\cdot\chi +\frac{t-1}{s-1-t}\nabla {\rm tr}\,\chi=0.\label{dedondergenstchi}\ee
Next consider the trace of the Fronsdal tensor, ${\rm tr}\,{\cal F}$.  Using \eqref{dedondergenst} to eliminate divergences of $\ell$ and then \eqref{trellchigen} to eliminate traces of $\ell$, followed by \eqref{dedondergenstchi} to eliminate divergences of $\chi$, we find,
\be {\rm tr}\,{\cal F}\propto  \rd_1^{(s-2,t-2)}{\rm tr}\,\chi.\ee
Given \eqref{Fvanishesgen} and the trivial cohomology assumption, we thus have ${\rm tr} \,\chi=0$, and hence from \eqref{dedondergenstchi} $\nabla\cdot\chi=0$.  We can therefore reach the gauge ${\rm tr}\, \ell=0$, and then by \eqref{dedondergenst} $\nabla\cdot\ell=0$.

Finally, inserting the conditions ${\rm tr}\,\ell=0$, $\nabla\cdot\ell=0$ into the Fronsdal-like tensor recovers the equations of motion \eqref{spinsdepthteom},
which are invariant under a residual gauge invariance reproducing \eqref{pmintogt}, \eqref{genresga}.
It can then be checked that the remaining equation of motion $\rd\ast{\cal K}= 0$ does not give any additional information, but is implied by the other equations, and that the field strength ${\cal K}$ is fully traceless.

\subsection{$d+1$ decomposition}

We can also understand the presence of duality in the general case by breaking up the field strength into space and time components, analogous to breaking up the Maxwell field strength into electric and magnetic fields.  Upon reducing $D\rightarrow d+1$, the fully traceless on-shell field strength tensor decomposes as
\bea
\begin{array}{|c c c c c|}\hline
&s\!\!\!\!\!\!&\!\!\!&&\\
\hline
\!\!\!  &\!\!\! t+1~~  \vline\!\!\!\\
\cline{1-2}
\end{array} ^{\ T}\!\!\!
\xrightarrow[{\scriptscriptstyle D\rightarrow d+1}]{}\hspace{-.5cm}
&& \begin{array}{|c c c c c|}\hline
&s\!\!\!\!\!\!&\!\!\!&&\\
\hline
\!\!\!  &\!\!\! t+1~~  \vline\!\!\!\\
\cline{1-2}
\end{array} ^{\ T}
~~\medoplus~~
\begin{array}{|c c c c|}\hline
&s-1\!\!\!\!\!\!&\!&\\
\hline
\!\!\!  &\!\!\! t+1~~  \vline\!\!\!\\
\cline{1-2}
\end{array} ^{\ T}
~~\medoplus~\cdots~\medoplus~~
\begin{array}{|c c c c|}\hline
\!\!\!  &\!\!\! ~t+1~~  \vline\!\!\!\\
\hline
\!\!\!  &\!\!\! ~t+1~~  \vline\!\!\!\\
\cline{1-2}
\end{array} ^{\ T}
 \nn \\
&&
\medoplus~~
\begin{array}{|c c c c c|}\hline
~&~s~~~~\!\!\!\!\!\!&\!\!\!&&\\
\hline
\!\!\!  &\!\!\!\!\!~t~~~~ \vline\!\!\!\!\!\\
\cline{1-2}
\end{array} ^{\ T}
 ~~\medoplus~~ 
\begin{array}{|c c c c c|}\hline
&s-1\!\!\!\!\!\!&\!\!\!&&\\
\hline
\!\!\!  &\!\! t~~~  \vline\!\!\!\!\!\!\\
\cline{1-2}
\end{array} ^{\ T}
~~\medoplus~\cdots~\medoplus ~~
\begin{array}{|c c c c c|}\hline
\!\!&t+1~\!\!\!\!\!\!\!\!&\!\!\!\!\!\!\!\!\!&&\\
\hline
\!\!\!  &\!\!\!\!\!~~t~~~ \vline\!\!\!\!\\
\cline{1-2}
\end{array} ^{\ T}
 \nn\\
&&
~~\!\vdots  \nn\\
&&
\medoplus~~\,
\begin{array}{|c c c c c|}\hline
&&s&&\\
\hline
~~~\!\!~\vline\!\!\!\\
\cline{1-1}
\end{array}  ^{\ T}
~~\medoplus~~
\begin{array}{|c c  c |}\hline
&\!\!s-1&\\
\hline
~~~\!\!~\vline\!\!\!\\
\cline{1-1}
\end{array} ^{\ T}
~~\medoplus~\cdots~\medoplus~~
\begin{array}{|c c  c |}\hline
&\!\!\!\!\!t+1\!\!\!\!&\\
\hline
~~~\!\!~\vline\!\!\!\\
\cline{1-1}
\end{array} ^{\ T}
\nn\\ 
&& 
\medoplus~~  
\begin{array}{|c c c c c|}\hline
&~&s&~&\\
\hline
\end{array}  ^{\ T}
~~\medoplus
~~ \begin{array}{|c c c|}\hline
&s-1&~\\
\hline
\end{array} ^{\ T}
~~\medoplus~\cdots~\medoplus~~
\!\begin{array}{|c c  c |}\hline
&\!\!\!t+1\!\!&\\
\hline
\end{array}  ^{\ T}\label{Ddbreakuptt}
\eea

In general $D$ we now have $t+2$ kinds of spatial field in the various lines of the right hand side of \eqref{Ddbreakuptt}.  In $D=4$, the representations in all but the final two lines carry no independent components and hence are not present.  The representations in the second to last line 
 can be dualized by hitting the anti-symmetric index pair with the spatial epsilon symbol $\epsilon_{ijk}$, after which they become ordinary symmetric tensors.  This leaves
 \bea \begin{array}{|c c c c c|}\hline
&&s&&\\
\hline
~~~\!\!~\vline\!\!\!\\
\cline{1-1}
\end{array}  ^{\ T}
\xrightarrow[{\scriptscriptstyle 4\rightarrow 3}]{}\hspace{-.5cm}
&& 
\begin{array}{|c c c c c|}\hline
&~&s&~&\\
\hline
\end{array}  ^{\ T}
~~\medoplus
~~ \begin{array}{|c c c|}\hline
&s-1&~\\
\hline
\end{array} ^{\ T}
~~\medoplus~\cdots~\medoplus~~
\!\begin{array}{|c c  c |}\hline
&\!\!\!t+1\!\!&\\
\hline
\end{array} ^{\ T}
 \\ 
&&\medoplus~~  
\begin{array}{|c c c c c|}\hline
&~&s&~&\\
\hline
\end{array}  ^{\ T}
~~\medoplus
~~ \begin{array}{|c c c|}\hline
&s-1&~\\
\hline
\end{array} ^{\ T}
~~\medoplus~\cdots~\medoplus~~
\!\begin{array}{|c c  c |}\hline
&\!\!\!t+1\!\!&\\
\hline
\end{array} ^{\ T} \, , \ \ \  D=4. \nn \label{Ddbreakuptt3}
\eea
The two lines now carry the same representation, and the action of duality is to rotate them into each other.  One can think of these fields as carrying the physical helicity components and canonical momenta for the helicities $s,s-1,\cdots,t+1$ of the depth $t$ PM field.

\subsection{Example: $s=4$, $t=2$\label{s4t2section}}
In order to make the previous discussion more concrete, here we work out the details for the simplest nontrivial case, depth $t=2$ spin-4. In this case, the fundamental object is a 7-index tensor
\be
{\cal K}_{\mu_1\nu_1|\mu_2\nu_2,\mu_3\nu_3, \alpha}\in~ 
\begin{ytableau}
 ~\\~
\end{ytableau}
~~\medotimes~~
\begin{ytableau}
 ~& ~ &~\\
~&~
\end{ytableau}\,. \label{s4t2rep}
\ee
\paragraph{Bianchi identities:}
As usual, we begin by considering the constraints that the Bianchi identities~\eqref{pmssbianchi} place on the form of the tensor \eqref{s4t2rep}. 
Decomposing the representation \eqref{s4t2rep}, we obtain the components 
\be
\begin{ytableau}
 ~\\~
\end{ytableau}
~~\medotimes~~
\begin{ytableau}
 ~& ~ &~\\
~&~
\end{ytableau}
~~=~~
\begin{ytableau}
 ~& ~ &~&~\\
~&~\\
~
\end{ytableau}
~~\medoplus~~
\begin{ytableau}
 ~& ~ &~\\
~&~&~\\
~
\end{ytableau}
~~\medoplus~~
\begin{ytableau}
 ~& ~ &~\\
~&~\\
~\\
~
\end{ytableau}
~~\medoplus~~
\begin{ytableau}
 ~& ~ &~&~\\
~&~&
\end{ytableau}\,.
\label{s4t2curv}
\ee
The algebraic Bianchi identity ${\rm tr}\ast {\cal K}=0$ gives 
\be
\epsilon_{\nu_1\beta_1\beta_2\cdots\beta_{D-2}}^{~~~~~~~~~~~~~~~~~\lambda\rho\sigma}{\cal K}_{\lambda\rho|\sigma\nu_2,\mu_{3}\nu_{3},\alpha}  = 0 \ \ \Rightarrow \ \ {\cal K}_{[\mu_1\nu_1|\mu_2]\nu_2,\mu_3\nu_3,\alpha} = 0
\ee
which means that every component in~\eqref{s4t2curv} vanishes except for the piece with two rows, 
\be
{\cal K}_{\mu_1\nu_1,\mu_2\nu_2,\mu_3\nu_3, \alpha}\in~\begin{ytableau}
 ~& ~ &~&~\\
~&~&
\end{ytableau}~.
\label{s4t2curvshape}
\ee

The differential Bianchi identity $\rd{\cal K} = 0$ fits into the complex \eqref{depthtspinscomplex} with $s=4$, $t=2$,%
\be
\begin{ytableau}
~&~
\end{ytableau}
\xrightarrow{
\overset{\rd^{(4,2)}_1}{\hspace*{.75cm}}}~
\begin{ytableau}
 ~& ~ &~&~\\
\end{ytableau}~
\xrightarrow{
\overset{\rd^{(4,2)}_2}{\hspace*{.75cm}}}~
\begin{ytableau}
 ~& ~ &~&~\\
~&~&
\end{ytableau}~
\xrightarrow{
\overset{\rd^{(4,2)}_3}{\hspace*{.75cm}}}~
\begin{ytableau}
 ~& ~ &~&~\\
~&~&\\
~
\end{ytableau}
\xrightarrow{\hspace*{.75cm}}\cdots
\ee
where the differential operators are given explicitly by
\begin{align}
\label{t2s4gauge}
\left(\rd^{(4,2)}_1\xi\right)_{\mu_1\mu_2\mu_3\mu_4} &= \nabla_{(\mu_1}\nabla_{\mu_2}\xi_{\mu_3\mu_4)} +H^2 g_{(\mu_1\mu_2}\xi_{\mu_3\mu_4)}\\\label{curvdiff}
\left(\rd^{(4,2)}_2\ell\right)_{\mu_1\nu_1,\mu_2\nu_2,\mu_3\nu_3, \alpha} &= P_{4,3}\left(\nabla_{\nu_1}\nabla_{\nu_2}\nabla_{\nu_3}+ 4H^2g_{\nu_1\nu_2}\nabla_{\nu_3}\right)\ell_{\mu_1\mu_2\mu_3\alpha}\\
\left(\rd^{(4,2)}_3 {\cal K}\right)_{\mu_1\nu_1\rho,\mu_2\nu_2,\mu_3\nu_3, \alpha} & = \nabla_{[\rho}{\cal K}_{\mu_1\nu_1],\mu_2\nu_2,\mu_3\nu_3, \alpha}\,.
\end{align}
and satisfy
\be \rd^{(4,2)}_{i+1}\circ\rd^{(4,2)}_{i}=0, \ \  i=1,2,\cdots.
\ee
Using the trivial cohomology assumption, we conclude that the Bianchi identity $\rd^{(4,2)}_3{\cal K} = 0$ allows us to write,
\be
{\cal K}_{\mu_1\nu_1,\mu_2\nu_2,\mu_3\nu_3, \alpha} = \left(\rd^{(4,2)}_{2}\ell\right)_{\mu_1\nu_1,\mu_2\nu_2,\mu_3\nu_3 ,\alpha}=P_{4,3}\left(\nabla_{\nu_1}\nabla_{\nu_2}\nabla_{\nu_3}+ 4H^2g_{\nu_1\nu_2}\nabla_{\nu_3}\right)\ell_{\mu_1\mu_2\mu_3\alpha},
\ee
for some rank-4 symmetric tensor $\ell$.  By virtue of $\rd^{(4,2)}_{2}\circ\rd^{(4,2)}_{1}=0$, ${\cal K}$ is gauge invariant under the gauge transformation
\be
\delta \ell_{\mu_1\mu_2\mu_3\alpha_1} = \left(\rd^{(4,2)}_1\xi\right)_{\mu_1\mu_2\mu_3\mu_4} = \nabla_{(\mu_1}\nabla_{\mu_2}\xi_{\mu_3\mu_4)} +H^2 g_{(\mu_1\mu_2}\xi_{\mu_3\mu_4)},
\label{s4t2gaugetrans}
\ee
for some two index fully symmetric gauge parameter $\xi_{\mu_1\mu_2}$.
\paragraph{Field equations:}
Start with the algebraic equation of motion ${\rm tr} \,{\cal K} = 0$.  This trace is a 5-index tensor which contains the symmetry components\footnote{Note that here ${\rm tr}~{\cal K}$ only has two symmetry components as opposed to the general case \eqref{trksymt2genc}.}
\be
{\rm tr} \,{\cal K}~\supset~
\begin{ytableau}
 ~& ~ &~&~\\
~
\end{ytableau}
~~\medoplus~~
\begin{ytableau}
 ~& ~ &~\\
~&~
\end{ytableau}\, .  \label{42tableauxtrKcomp}
\ee
If we project  onto the first tableau, using indices~
\scalebox{.55}{$
\begin{ytableau}
\nu_1 & \nu_2 & \mu_3 &\alpha \\
\nu_3
\end{ytableau}
$},
we find that we can write the result as $\rd^{(4,0)}_2$ of a 4-index tensor ${\cal F}$,
\be
P_{4,1}{\cal K}^{\lambda}_{~~\nu_1,\lambda\nu_2,\mu_3\nu_3,\alpha} \propto \rd^{(4,0)}_2{\cal F} =2 \partial_{[\nu_1}{\cal F}_{\nu_3]\nu_2\mu_3\alpha}~,
\ee
where
\be
{\cal F}_{\mu_1\mu_2\mu_3\mu_4} = \square \ell_{\mu_1\mu_2\mu_3\mu_4}+2 \nabla_{(\mu_1}\nabla_{\mu_2}\ell_{\mu_3\mu_4)\nu}^{~~~~~~~\nu}-\frac{8}{3}\nabla_{(\mu_1}\nabla_{|\nu|} \ell^\nu_{~~\mu_2\mu_3\mu_4)}+ H^2(D-4)\ell_{\mu_1\mu_2\mu_3\mu_4}+6 H^2 g_{(\mu_1\mu_2}\ell_{\mu_3\mu_4)\nu}^{~~~~~~~\nu}\,.
\label{s4t2frons}
\ee
The tensor ${\cal F}$ is totally symmetric and is the PM analogue of the Fronsdal tensor.
The equation ${\rm tr}\,{\cal K} = 0$ therefore implies that $\rd^{(4,0)}_2{\cal F}=0$.
This implies, using the $t=0$ complex \eqref{spinst0complex} where $s=4$, that ${\cal F}$ can be written as gradients acting on some scalar $\chi$, as in \eqref{t0gauge},
\be
 {\cal F}_{\mu_1\mu_2\mu_3\mu_4}  = \rd^{(4,0)}_1\chi=\left(\nabla_{(\mu_1}\nabla_{\mu_2}\nabla_{\mu_3}\nabla_{\mu_4)}+10H^2 g_{(\mu_1\mu_2}\nabla_{\mu_3}\nabla_{\mu_4)}+9H^4 g_{(\mu_1\mu_2}g_{\mu_3\mu_4)}\right)\chi~.\label{fchi42form}
\ee
Next, we note that under a gauge transformation~\eqref{s4t2gaugetrans}, ${\cal F}$ transforms as
\bea
\delta {\cal F}_{\mu_1\mu_2\mu_3\mu_4}  = \left(\nabla_{(\mu_1}\nabla_{\mu_2}\nabla_{\mu_3}\nabla_{\mu_4)}+10H^2 g_{(\mu_1\mu_2}\nabla_{\mu_3}\nabla_{\mu_4)}+9H^4 g_{(\mu_1\mu_2}g_{\mu_3\mu_4)}\right)\xi^\nu_{~\nu}\,=\left(\rd^{(4,0)}_1{\rm tr} \,\xi \right)_{\mu_1\mu_2\mu_3\mu_4}, \nn\\
\eea
which is the depth $t=0$ gauge transformation with scalar gauge parameter $\xi_{~\beta}^\beta$, exactly of the form \eqref{fchi42form}.  Using the trivial cohomology assumption of the complex \eqref{spinst0complex}, this implies that we can gauge-fix ${\cal F}$ to zero by using the trace of the gauge parameter, ${\rm tr }\,\xi$.
After fixing ${\cal F}=0$ we have a residual gauge freedom given by any traceless gauge parameter, ${\rm tr}\,\xi=0$. 

Next we project the ${\rm tr}\,{\cal K} = 0$ equation onto the second factor of \eqref{42tableauxtrKcomp} using the tableau \scalebox{.55}{$
\begin{ytableau}
 \nu_1 & \nu_2  &\nu_3\\
\mu_3 &\alpha 
\end{ytableau}
$}
.  We get (after renaming some indices)
\begin{align}
P_{3,2}\left(\nabla_{\nu_1}\nabla_{\nu_2}+H^2g_{\nu_1\nu_2}\right)\Big[\nabla_\rho \ell^{\rho}_{~\mu_1\mu_2\mu_3}-3\nabla_{(\mu_1}\ell^{~~~~~~\rho}_{\mu_2\mu_3)\rho}
\Big]=0\,.
\end{align}
This equation is nothing but $\rd^{(3,1)}_2$ acting on a de Donder-like condition for the gauge field $\ell$.
using the complex~\eqref{depthtspinscomplex}, this implies that we can write 
\be \nabla_\rho \ell^{\rho}_{~\mu_1\mu_2\mu_3}-3\nabla_{(\mu_1}\ell^{~~~~~~\rho}_{\mu_2\mu_3)\rho} = \left(\rd_1^{(3,1)}\chi\right)_{\mu_1\mu_2\mu_3}\, ,\label{dedchis4t2}\ee
with $\chi_\mu$ some vector parameter.  Under a gauge transformation, the de Donder-like combination which is the left hand side of ~\eqref{dedchis4t2} transforms as
\be
\delta\left(\nabla_\rho \ell^{\rho}_{~\mu_1\mu_2\mu_3}-3\nabla_{(\mu_1}\ell^{~~~~~~\rho}_{\mu_2\mu_3)\rho}\right) = -\frac{3}{2} \left(\nabla_{(\mu_1}\nabla_{\mu_2}+H^2 g_{(\mu_1\mu_2}\right)\nabla^\rho \xi_{\mu_3)\rho}\propto\left(\rd_1^{(3,1)}\nabla\cdot\xi \right)_{~\mu_1\mu_2\mu_3}  .
\ee
Comparing this to \eqref{dedchis4t2}, we see that we can fix the de Donder type gauge
\be
\nabla_\rho \ell^{\rho}_{~\mu_1\mu_2\mu_3}-3\nabla_{(\mu_1}\ell^{~~~~~~\rho}_{\mu_2\mu_3)\rho} = 0\,,
\label{eq:dedondergauges4t2}
\ee
by solving the equation $\chi \propto\nabla\cdot\xi = 0$ for $\xi$.  This leaves a residual gauge freedom satisfying $\nabla\cdot\xi = 0$. 

Consider now the gauge transformation of the trace of $\ell$.  Using the fact that the gauge parameter has been fixed to be transverse and traceless, we have
\be
\delta\ell_{\mu_1\mu_2\rho}^{~~~~~~\rho} = \frac{1}{6}\left(\square+(4+3D)H^2\right)\xi_{\mu_1\mu_2}.
\ee
We want to reach the gauge ${\rm tr}\,\ell=0$, and to do this with a transverse traceless $\xi_{\mu\nu}$, we need to argue that ${\rm tr}\,\ell=0$ is itself transverse and traceless, after which we will be able to reach ${\rm tr}\,\ell=0$ leaving a residual transverse traceless gauge parameter satisfying the homogeneous equation $\left(\square+(4+3D)H^2\right)\xi_{\mu_1\mu_2}=0$.

In fact, the trace of the gauge field is transverse and traceless.  Taking a trace of the de Donder condition \eqref{eq:dedondergauges4t2} gives us a de Donder-like condition for ${\rm tr}\,\ell$,
\be \nabla _{\nu}\ell_{\  \mu\  \rho }^{\nu\  \rho}=-\nabla_\mu \ell^{\nu\  \rho}_{\  \nu \ \rho}=0.\label{dedondertrl}\ee
 Taking a trace of the Fronsdal tensor \eqref{s4t2frons}, and then using the de Donder condition \eqref{eq:dedondergauges4t2} and its trace \eqref{dedondertrl} to eliminate all divergences, we find that it becomes
\be {\cal F}_{\mu_1\mu_2 \ \nu}^{\ \ \ \ \  \nu}=3\left(\nabla_{\mu_2}\nabla_{\mu_2}+H^2g_{\mu_1\mu_2}\right)\ell^{\nu\  \rho}_{\  \nu \ \rho} .\ee
The right hand side is nothing but $\rd_1^{(2,0)}{\rm tr}^2\,\ell$, so the vanishing of the Fronsdal tensor, along with the trivial cohomology assumption for the complex with $s=2$, $t=0$ tells us that the double trace ${\rm tr}^2\ell$ vanishes,
\be \ell^{\nu\  \rho}_{\  \nu \ \rho}=0\, \label{vanishingdoubtr42},\ee
after which \eqref{dedondertrl}  tells us that ${\rm tr}\,\ell$ is transverse.  We can therefore reach the gauge  ${\rm tr}\, \ell=0$, after which \eqref{eq:dedondergauges4t2} tells us that $\ell$ is transverse.

Inserting ${\rm tr}\, \ell=0$ and $\nabla\cdot\ell = 0$ into ${\cal F} = 0$, we obtain the system of equations
\be\label{s4t2finaleql}
\left(\square + (D-4)H^2\right)\ell_{\mu_1\mu_2\mu_3\mu_4} = 0~,~~~~~~~~~~\nabla^\nu \ell_{\nu\mu_2\mu_3\mu_4} = 0~,~~~~~~~~~~\ell^\nu_{\ \nu\mu_3\mu_4} = 0~,
\ee
which are the correct on shell equations \eqref{spinsdepthteom} for a PM spin-4 of depth-2. These equations have a residual gauge symmetry where
\be
\delta\ell_{\mu_1\mu_2\mu_3\mu_4} = \nabla_{(\mu_1}\nabla_{\mu_2}\xi_{\mu_3\mu_4)} +H^2 g_{(\mu_1\mu_2}\xi_{\mu_3\mu_4)}\,,~~\xi^\nu_{~\,\nu} = 0\,,~~\nabla_\nu\xi^\nu_{\,~\mu}=0\,,~~\left(\square+(4+3D)H^2\right)\xi_{\mu_1\mu_2}=0,
\ee
which are the correct on shell gauge symmetries \eqref{pmintogt} for a PM spin-4 of depth-2.
Finally, it is straightforward to show that upon using \eqref{s4t2finaleql} the remaining equation of motion $\rd\ast{\cal K}= 0$ is automatically satisfied, and that the field strength ${\cal K}$ is fully traceless.

\section{Conclusions}
\label{sec:conclusions}

We have seen how the equations of motion for integer spin partially massless fields can be recovered from imposing equations on gauge-invariant curvature tensors. The benefit of this formulation is that it allows us to see the $D=4$ electric-magnetic-like duality of these theories in a manifestly local and de Sitter covariant form.  However, it should be noted that writing the equations of motion in duality covariant form, though suggestive, does not by itself establish duality invariance of the action.  The 3+1 formulation of \cite{Deser:2013xb}, on the other hand, shows invariance of the action, at the unavoidable price of losing manifest de Sitter invariance.  

Writing the equations in duality covariant form complements the 3+1 analysis, and in particular paves the way for us to introduce magnetic sources, which cannot be introduced locally into the action. For massless and massive higher spins, monopole solutions were constructed in~\cite{Nepomechie:1984wu,Teitelboim:1985ya,Teitelboim:1985yc,Bunster:2006rt}. It would interesting to see if the electric and magnetic monopole solutions of the PM spin-2 case \cite{Hinterbichler:2015nua} generalize to the higher spins.

There are various extensions and generalizations which naturally present themselves.
It would be interesting to construct gauge invariant actions for higher spin and depth partially massless fields utilizing these metric-like curvature tensors or Fronsdal tensors. Such a construction is known for the spin-2 case~\cite{Deser:2006zx}, but it is likely that the higher-spin analogues will require introduction of auxiliary fields, as the known actions for massive higher spins require such auxiliary fields. In~\cite{Skvortsov:2006at}, Skvortsov and Vasiliev give a construction of actions for similar curvature tensors in the frame-like formulation, which from the point of view of the metric formulation contains many auxiliary and St\"uckelberg-like fields. More generally, it would be interesting to further elucidate the relationship between the curvature tensors constructed here and those of~\cite{Skvortsov:2006at}. In particular, the EM duality should act in a very simple way in the frame-like formulation, by dualizing the form indices of the curvature tensors.

We have considered here only bosonic higher spin fields. There appears to be no obstruction to constructing similar gauge invariant curvatures for fermionic partially massless fields \cite{Deser:1983tm,Deser:1983mm,Deser:2001xr,Deser:2001us,Deser:2001pe,Deser:2014ssa}, or for mixed symmetry fields~\cite{Curtright:1980yk} which can also possess partially massless points~\cite{Brink:2000ag,Zinoviev:2002ye,Zinoviev:2003dd,Skvortsov:2009zu}. Some technical issues that might be interesting to investigate include proving curved space versions of the generalized Poincar\'e lemmas of~\cite{DuboisViolette:1999rd,DuboisViolette:2001jk,Bekaert:2002dt}. Additionally, we have not provided an explicit construction of the PM Fronsdal-type \cite{Fronsdal:1978rb} tensors for all depths and spins -- an explicit expression for these tensors may shed some light onto the problem of constructing explicit actions from these curvature tensors.

Moving beyond the linear case, one intriguing possible application is to attempt to construct non-Abelian theories of partially massless fields. Similar to the construction of Yang--Mills \cite{Yang:1954ek} or the Fradkin--Vasiliev procedure \cite{Fradkin:1986qy,Fradkin:1987ks} (see e.g.,~\cite{Bekaert:2005vh,Didenko:2014dwa}), it might be possible to construct generalizations of these linear field strengths which are invariant under non-linear symmetries. Some work along these directions appears in~\cite{Boulanger:2011se,Boulanger:2012dx,Joung:2012rv,Joung:2012hz}. In the spin-2 case, this does not appear to be possible~\cite{Garcia-Saenz:2015mqi}, but the question for higher spins remains open.

\vspace{-.5cm}
\paragraph{Acknowledgements:} We thank Garrett Goon, Rachel Rosen, Andrew Waldron and George Zahariade for helpful discussions.  We would like to thank the Sitka Sound Science Center for hospitality while some of this work was completed. This work was supported in part by National Science Foundation Grant No. PHYS-1066293 and the hospitality of the Aspen Center for Physics. This work was also supported in part by the Kavli Institute for Cosmological Physics at the University of Chicago through grant NSF PHY-1125897, an endowment from the Kavli Foundation and its founder Fred Kavli, and by the Robert R. McCormick Postdoctoral Fellowship (AJ).

\appendix

\section{Off shell spin-3}
\label{spin3workedout}
Here we consider the full off-shell Lagrangian for a spin-3 field propagating on a maximally symmetric space of nonzero curvature, showing how the on-shell equations of motion \eqref{massivefields} and the partially massless points \eqref{pmpoints} arise. 
The equations of motion and gauge invariances for this case are also studied in \cite{Deser:2001pe,Deser:2001us,Gover:2014vxa}.
Spin-3 is the simplest example where the field exhibits multiple partially massless depths in addition to the ordinary massless point. The Lagrangian for a massive spin-3 propagating on a maximally-symmetric space is\footnote{This Lagrangian can be obtained by performing a radial dimensional reduction of a massless spin-3 field in $(D+1)$-dimensions \cite{Biswas:2002nk,Hallowell:2005np}.}
\begin{align} 
{\cal L}=&-{1\over 2}\nabla_\mu b_{\nu\alpha\beta} \nabla^\mu b^{\nu\alpha\beta}+{3\over 2}\nabla_\mu b^{\mu}_{\ \alpha\beta} \nabla_\nu b^{\nu\alpha\beta}-3\nabla_\mu b_{\nu\alpha}^{\ \ \alpha} \nabla_\beta b^{\beta \mu\nu} +{3\over 2} \nabla_\mu b_{\nu\alpha}^{\ \ \alpha}\nabla^\mu b^{\nu\beta}_{\ \ \beta}+{3\over 4}\nabla_\mu b^{\mu\alpha}_{\ \ \alpha} \nabla_\nu b^{\nu\beta}_{\ \ \beta} \nonumber\\\nonumber
&-{1\over 2}m^2\left(b_{\mu\nu\alpha}b^{\mu\nu\alpha}-3b_{\mu\alpha}^{\ \ \alpha}b^{\mu\beta}_{\ \ \beta}\right)-{3(D-2)\over 2D}m h\nabla_\mu b^{\mu\nu}_{\ \ \nu}+{3(D-2)(D-1)\over 2D^2}(\nabla h)^2+{9\over 4}m^2 h^2\\
&+\frac{(D-1)H^2}{2}\left(\frac{D-3}{D-1}b_{\mu\nu\rho}b^{\mu\nu\rho} - 6 b_{\mu\alpha}^{~~\alpha}b^{\mu\beta}_{~~\beta}-9h^2\right)~.
\label{spin3curvedspacelag}
\end{align}
The dynamical variables are the symmetric, trace-ful, spin 3 field $b_{\mu\nu\rho}$, and a scalar field, $h$, which will end up being non-dynamical but is necessary in order to obtain the correct degrees of freedom for a spin 3.

This Lagrangian possesses 3 special values of the mass, $m$. The first is $m=0$, corresponding to a massless spin-3 ($t=2)$.  In this case, the scalar field $h$ decouples and we acquire the Fronsdal gauge symmetry
\be
\delta b_{\mu\nu\rho} = \nabla_{(\mu}\Lambda_{\nu\rho)} \, ,
\label{freegauges}
\ee
where the gauge parameter is symmetric and traceless: $\Lambda_{[\mu\nu]}=  \Lambda_\mu^\mu = 0$.  This symmetry removes the helicity-0, helicity-1 and helicity-2 polarizations, leaving only the massless helicity-3.

The next special mass value is the $t=1$ partially massless point,
\be
m^2 = D H^2 \, . 
\ee
At this point the field has a vector gauge symmetry\footnote{Note that in the AdS case where $H^2\rightarrow -{1\over L^2}$ (with $L$ the AdS radius) the square roots in \eqref{eq2s3dxi}, \eqref{eq2s3dxi2} become imaginary.  In this case we must also replace $h\rightarrow ih$, which keeps the Lagrangian real.} with gauge parameter $\xi_\mu$,
\begin{align}
\delta b_{\mu\nu\rho} &= \nabla_{(\mu}\nabla_{\nu}\xi_{\rho)} - \frac{1}{D}g_{(\mu\nu}\nabla_{\rho)}\nabla_\alpha\xi^\alpha+ H^2 g_{(\mu\nu}\xi_{\rho)}\, ,\nn \\
\delta h &= -\frac{1}{3} \sqrt{DH^2} \nabla_\mu\xi^\mu~.\label{eq2s3dxi}
\end{align}
This symmetry removes both the helicity-0 and helicity-1 polarizations.

The final special mass value is the $t=0$ partially massless point
\be
m^2 = 2(D-1)H^2\, . 
\ee
At this value of the mass, there is a scalar gauge symmetry with gauge parameter $\chi$, which acts as
\begin{align}
\label{PMspin3symm}
\delta b_{\mu\nu\rho} &= \nabla_{(\mu}\nabla_\nu\nabla_{\rho)}\chi - \frac{1}{D}g_{(\mu\nu}\nabla_{\rho)}\square\chi+\frac{2(D-1)H^2}{D} g_{(\mu\nu}\nabla_{\rho)}\chi\, ,\\
\delta h & = -\frac{1}{3}\sqrt{2(D-1)H^2}\Big(\square +2(D+1)H^2\Big)\chi~, \label{eq2s3dxi2}
\end{align}
and removes the helicity-0 polarization.

\subsection{St\"uckelberg and decoupling limit}
An elegant way to understand the nature of these partially massless points is to employ the St\"uckelberg trick.  The gauge symmetry~\eqref{freegauges} of the massless theory is broken by the presence of the mass terms in~\eqref{spin3curvedspacelag}, but can be restored by introducing St\"uckelberg fields $h_{\mu\nu}$, $A_\mu$, $\phi$ (with $h_{\mu\nu}$ symmetric and trace-ful), associated to the helicity $2, 1$ and $0$ components respectively, through the replacement\footnote{In this replacement, the auxiliary scalar field becomes the trace of the St\"uckelberg field $h_{\mu\nu}$, i.e., what we are doing is to introduce a traceless symmetric tensor field as a St\"uckelberg and then package the original auxiliary field as the trace of this tensor field.  This St\"uckelberg replacement emerges naturally from dimensional reduction, see e.g., \cite{Porrati:2008ha}.}  
\begin{align}
h &\longmapsto mh +\frac{m^3}{3}\phi+m\nabla_\mu A^\mu-\frac{m}{3}\square\phi\\
b_{\mu\nu\rho}&\longmapsto b_{\mu\nu\rho}-3\nabla_{(\mu}h_{\nu\rho)}-\frac{3}{D}\eta_{(\mu\nu}\nabla_{\rho)}h+3\nabla_{(\mu}\nabla_{(\nu}A_{\rho))}\\\nonumber&~~~~~~-\frac{3}{D}\eta_{(\mu\nu}\nabla_{\rho)}\nabla_\alpha A^\alpha-\nabla_{(\mu}\nabla_\nu\nabla_{\rho)}\phi - \frac{1}{D }\eta_{(\mu\nu}\nabla_{\rho)}\square\phi +\frac{m^2}{D  }\eta_{(\mu\nu}\nabla_{\rho)}\phi~.
\end{align}

In order to isolate the individual helicity components, we take the decoupling limit
\be
m\to 0\, ,~~~~~~~~~~H\to0\, ,~~~~~~~~~~\frac{m}{H}\to~{\rm fixed}\ \ \ .
\ee
Additionally, we perform the following two field redefinitions to diagonalize the $b,A$ and $h,\phi$ mixing:
\begin{align}
b_{\mu\nu\rho} &\longmapsto b'_{\mu\nu\rho} - \frac{3}{D}H^2g_{(\mu\nu}A_{\rho)}\, ,\\
h_{\mu\nu} &\longmapsto h'_{\mu\nu}  -m^2 \frac{1+\frac{2H^2}{3 m^2}(D+1)}{(D-2)}g_{\mu\nu}\phi~,
\end{align}
and we keep the canonically normalized fields 
\be \hat b\sim b,~~~~~ \hat h\sim { m}h,~~~~~ \hat A\sim { m^2}A,~~~~~ \hat \phi\sim { m^3}\phi,\ee
 fixed as we take the limit.
After all of these manipulations, the Lagrangian~\eqref{spin3curvedspacelag} becomes a flat space Lagrangian and takes the following form
\begin{align}
\nonumber
{\cal L} = &-{1\over 2}\partial_\mu b'_{\nu\alpha\beta} \partial^\mu b'^{\nu\alpha\beta}+{3\over 2}\partial_\mu b'^{\mu}_{\ \alpha\beta} \partial_\nu b'^{\nu\alpha\beta}-3\partial_\mu {b'}_{\nu\alpha}^{\ \ \alpha} \partial_\beta b'^{\beta \mu\nu} +{3\over 2} \partial_\mu {b'}_{\nu\alpha}^{\ \ \alpha}\partial^\mu {b'}^{\nu\beta}_{\ \ \beta}+{3\over 4}\partial_\mu {b'}^{\mu\alpha}_{\ \ \alpha} \partial_\nu {b'}^{\nu\beta}_{\ \ \beta}\\ \nonumber
&+3m^2\left(-\frac{1}{2}\partial_\lambda h'_{\mu\nu}\partial^\lambda h'^{\mu\nu}+\partial_\mu h'_{\nu\lambda}\partial^\nu h'^{\mu\lambda}-\partial_\mu h'^{\mu\nu}\partial_\nu h'+\frac{1}{2}\partial_\lambda h'\partial^\lambda h'\right)\\ \nonumber
&-\frac{3(D+1)}{4D} m^2(m^2-DH^2)\left(\partial_\mu A_\nu-\partial_\nu A_\mu\right)^2\\
&+\frac{(D+1)}{2(D-2)}m^2(m^2-DH^2)(m^2-2(D-1)H^2)\phi\square\phi~, \label{decoupls3actn}
\end{align}
and is invariant under the linear gauge symmetries:
\bea \delta b'_{\mu\nu\rho}&=&\partial_\mu\lambda_{\nu\rho}^T+\partial_\nu\lambda_{\rho\mu}^T+\partial_\rho\lambda_{\mu\nu}^T\, ,\\
\delta h'_{\mu\nu}&=& \partial_\mu\lambda_\nu+\partial_\nu\lambda_\mu \, ,\\
\delta A_\mu&=& \partial_\mu\lambda\, ,
\eea
where $\lambda_{\mu\nu}^T$ is a symmetric, traceless gauge parameter.

\begin{figure}[h!]
\centering
~~~~~~~~~~~~~~~~~~~~~~~~~~~~~~~~
\epsfig{file=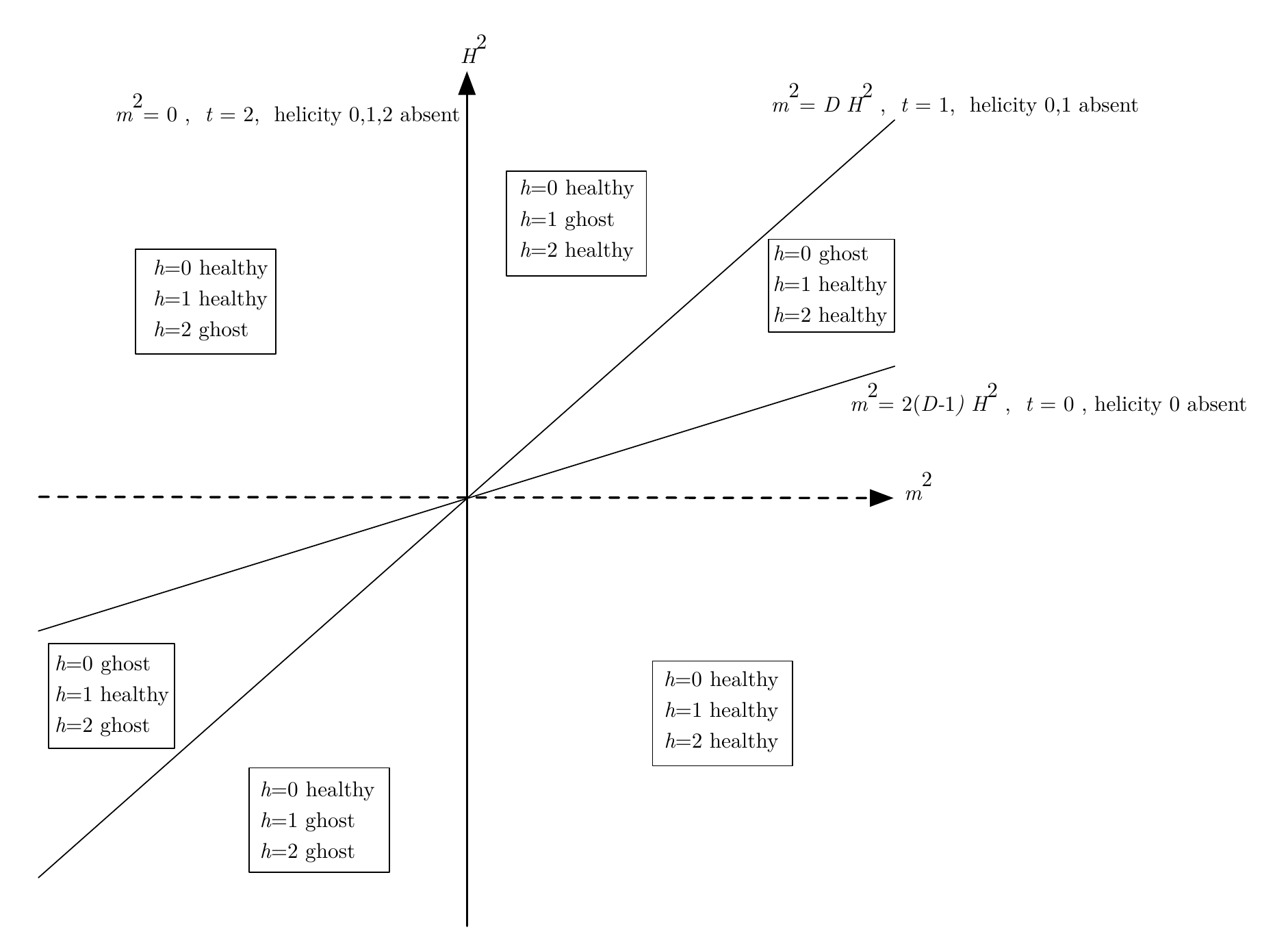,height=4in,width=5.5in}
\caption{\small Partially massless lines and regions for spin 3.  Modes pass between being healthy and ghostlike as the partially massless lines are crossed.}
\label{spin3fig}
\end{figure}

The action \eqref{decoupls3actn} is now a flat-space action for massless fields of helicity 3, 2, 1, 0, all decoupled from each other.  We see manifestly the three partially massless points; these are the points where kinetic terms for the various helicity components vanish.  Furthermore, for any value of $m\over H$ we can determine each component's unitarity/ghostliness from the sign in front of the kinetic terms.   At the depth $t=0$ line, $m^2=2(D-1)H^2$, the scalar degree of freedom drops out of the Lagrangian. Similarly, at the $t=1$ line $m^2=DH^2$, both the scalar and vector polarizations are removed, and at $m^2=0$ the helicity $0,1,2$ components are all removed. These points mark the boundaries between healthy and ghost-like regions for the various helicity components, see Figure \ref{spin3fig}.  In particular, we see immediately that on dS all components are healthy only for $m^2\geq 2(D-1)H^2$ (the Higuchi bound \cite{Higuchi:1986py} for spin 3), with the exception of the partially massless and massless points, which are also healthy.  On AdS, all components are healthy only for $m^2\geq 0$; the massless point is healthy and the partially massless points are ghost-like.  In general, for all spins $s\geq 1$, the partially massless cases, $t<s-1$, are healthy on dS and ghostly on AdS, whereas the massless cases, $t=s-1$, are healthy on both dS and AdS.

\subsection{Equations of motion}

Here we derive the equations of motion from the spin-3 Lagrangian \eqref{spin3curvedspacelag} and show that they reproduce the general on-shell equations~\eqref{massivefields}.  Denoting the Euler--Lagrange derivative of~\eqref{spin3curvedspacelag} with respect to $b_{\mu\nu\rho}$ as ${\cal E}_{\mu\nu\rho}$, it is convenient to work with the trace-reversed equations of motion
\begin{align}
\label{curlyGeq}
{\cal G}_{\mu\nu\rho} \equiv & ~{\cal E}_{\mu\nu\rho}-\frac{3}{D}\eta_{(\mu\nu}{\cal E}_{\rho)\alpha}^{~~~\alpha}\\\nonumber
=& \left(\square-m^2+4DH^2\right)b_{\mu\nu\rho} + 3\nabla_{(\mu}\nabla_{\nu}b_{\rho)\alpha}^{~~~\alpha} - 3\nabla^\alpha\nabla_{(\mu}b_{\nu\rho)\alpha}-\frac{3}{D}m^2g_{(\mu\nu}b_{\rho)\alpha}^{~~~\alpha}-\frac{3(D-2)}{D^2}mg_{(\mu\nu}\nabla_{\rho)}h = 0\, .
\end{align}
The trace-free divergence, which vanishes in the massless theory by the Bianchi identity associated to the gauge invariance, is proportional to the mass in the massive theory
\be
\nabla^\rho{\cal E}_{\rho(\mu\nu)_T} = m^2\left(2\nabla_{(\mu}b_{\nu)\alpha}^{~~~\alpha}-\nabla^\alpha b_{\alpha\mu\nu}-\frac{1}{D}g_{\mu\nu}\nabla_\alpha b^{\alpha\beta}_{~~\beta}\right)+\frac{(D-2)}{D}m\left(\nabla_\mu\nabla_\nu h-\frac{1}{D}g_{\mu\nu}\square h\right) = 0.
\label{tracefreedivergence}
\ee
We also have the $h$ equation of motion:
\be
{\cal H} \equiv {\delta {\cal L}\over \delta h}= \frac{9}{2}\Big(m^2 - 2(D-1)H^2\Big)h - \frac{3(D-2)}{2D}\left(\frac{2(D-1)}{D}\square h + m \nabla_\mu b^{\mu\alpha}_{~~~\alpha} \right)=0~.
\label{Heom}
\ee
Taking the following combination
\be
\label{GHEcombo}
\nabla^\nu\nabla^\rho{\cal E}_{\rho(\mu\nu)_T}-\frac{m^2}{2}{\cal G}_{\mu\alpha}^{~~~\alpha}+\frac{m}{3}\nabla_\mu{\cal H} = 0~,
\ee
yields the following constraint equation,
\be
\frac{(D+1)}{D}\left(m^2 - DH^2\right)\left(m^2 b_{\mu\nu}^{~~~\nu}+\frac{2(D-1)}{D}m\nabla_\mu h\right) = 0~,
\ee
which implies that (except when $m^2 = DH^2$)
\be
b_{\mu\nu}^{~~~\nu}=-\frac{2(D-1)}{mD}\nabla_\mu h~. \label{tbheqa}
\ee
Plugging this back into \eqref{Heom} we find
\be
\frac{9}{2}\Big(m^2 - 2(D-1)H^2\Big)h =0~,
\label{heq}
\ee
from which we deduce that (except when $m^2 = 2(D-1)H^2$)
\be
h = 0~, \label{htraceless}
\ee
and therefore, using \eqref{tbheqa},
\be
b_{\mu\nu}^{~~~\nu}=0. \label{btraceless}
\ee
Using \eqref{htraceless} and \eqref{btraceless} in~\eqref{tracefreedivergence} implies that (except when $m^2=0$)
\be
\nabla^\mu b_{\mu\nu\rho} = 0.
\ee
Now, putting this all together in ${\cal G}_{\mu\nu\rho}$ we obtain (note that we have to commute some $\nabla$s to use the divergence-free condition)
\be
\Big(\square-m^2+(D-3)H^2\Big)b_{\mu\nu\rho} = 0.
\ee
So we see that all together, the equations of motion following from~\eqref{spin3curvedspacelag} are
\be
\Big(\square-m^2+(D-3)H^2\Big)b_{\mu\nu\rho} = 0~,~~~~~~~\nabla^\mu b_{\mu\nu\rho} = 0~,~~~~~~~b_{\mu\nu}^{~~\nu} = 0~,~~~~~~~h=0~.
\ee
Note that this procedure breaks down at precisely the partially massless and massless points
\be
m^2 = DH^2~,~~~~m^2 = 2(D-1)H^2~,\ \ \ \ m^2=0\, .
\ee
We now proceed to treat the partially massless points separately.

\subsubsection{$t=0$ partially massless point}
At the partially massless point $m^2 = 2(D-1)H^2$, the procedure outlined above
does not quite work because~\eqref{heq} vanishes identically, which is nothing but the Noether identity associated with the partially massless gauge symmetry \eqref{PMspin3symm}. We can still combine~\eqref{curlyGeq},~\eqref{tracefreedivergence} and~\eqref{Heom} as in~\eqref{GHEcombo} to obtain a relation between $b_{\mu\nu}^{~~~\nu}$ and $\nabla_\mu h$:
\be
b_{\mu\nu}^{~~~\nu}=-\frac{\sqrt{2(D-1)}}{DH}\nabla_\mu h~, \label{t0bhconstr}
\ee
and plug this constraint back into \eqref{tracefreedivergence} and \eqref{curlyGeq} to obtain:
\begin{align}
\label{t0curlyE}
\nabla^\rho{\cal E}_{\rho(\mu\nu)_T} &= \frac{(3D-2)}{2(D-1)}\nabla_{(\mu}b_{\nu)\alpha}^{~~~\alpha}-\nabla^\alpha b_{\alpha\mu\nu}-\frac{1}{2(D-1)}g_{\mu\nu}\nabla^\alpha b_{\alpha\beta}^{~~~\beta} = 0\, , \\
{\cal G}_{\mu\nu\rho} &= \Big(\square+2(D+1)H^2\Big) b_{\mu\nu\rho}+3\nabla_{(\mu}\nabla_{\nu} b_{\rho)\alpha}^{~~~\alpha}-3\nabla^\alpha\nabla_{(\mu}b_{\nu\rho)\alpha} - 3H^2 g_{(\mu\nu}b_{\rho)\alpha}^{~~~\alpha} = 0\, .
\label{t0curlyG}
\end{align}
Now, we can use the PM gauge symmetry \eqref{eq2s3dxi2} to gauge fix $h=0$, which leaves a residual gauge symmetry satisfying 
\be
\left(\square + 2(D+1)H^2\right)\chi=0~. \label{chiresidt0}
\ee
In this gauge, we then have from \eqref{t0bhconstr} that $b$ is traceless,
\be b_{\mu\nu}^{~~~\nu} = 0,\ee
and 
\eqref{t0curlyE} then tells us that $b$ is divergenceless, 
\be \nabla^\rho b_{\rho\mu\nu} = 0,\ee
so we arrive at the following equations of motion
\be
\left(\square-(D+1)H^2\right)b_{\mu\nu\rho} = 0,~~~~~~~\nabla^\mu b_{\mu\nu\rho} = 0~,~~~~~~~b_{\mu\alpha}^{~~\alpha} = 0~,~~~~~~~h=0~,
\label{s3maximaldeptheoms}
\ee
which are invariant under the residual scalar gauge symmetry 
\be
\delta b_{\mu\nu\rho}= \nabla_{(\mu}\nabla_\nu\nabla_{\rho)}\chi +4H^2 g_{(\mu\nu}\nabla_{\rho)}\chi\,,~~{\rm with}~~~~~ \left(\square + 2(D+1)H^2\right)\chi=0. \label{PMspin3symmon}
\ee
Note that we have used \eqref{chiresidt0} in \eqref{PMspin3symm} to arrive at \eqref{PMspin3symmon}.  This recovers the on-shell equations \eqref{onshellexamt0} and gauge symmetries \eqref{s3t0gaugesymmex}.
\subsubsection{$t=1$ partially massless point}
\label{sec:t1lag}
At the partially massless point $m^2 = DH^2$, the theory possess the vector gauge invariance \eqref{eq2s3dxi}. We first use this gauge symmetry to gauge fix 
\be h=0\, ,\label{h0app31gauge}\ee
by solving the first order equation $h+\delta h=h -\frac{1}{3} \sqrt{DH^2} \nabla_\mu\xi^\mu=0$ for $\xi^\mu$.  This leaves a residual gauge symmetry which consists of any $\xi^\mu$ satisfying the homogeneous part of the equation, $\nabla_\mu\xi^\mu=0$.

 Taking the trace of the $b_{\mu\nu\rho}$ transformation \eqref{eq2s3dxi}, and using the residual gauge condition $\nabla_\mu\xi^\mu=0$, we learn that the trace transforms as
\be \delta b_{\mu\nu}^{~~~\nu} ={1\over 3}\left(\square + (1+2D)H^2\right)\xi_\mu  . \label{xiwaveeqap31}\ee
We want to reach a gauge where $b_{\mu\nu}^{~~~\nu} = 0$.  To do this, we must solve $b_{\mu\nu}^{~~~\nu} +\delta b_{\mu\nu}^{~~~\nu} =0$ for $\xi^\mu$ within the space of $\xi^\mu$ satisfying $\nabla_\mu\xi^\mu=0$.  This can be done, because taking a divergence, we find the source $\nabla^\mu b_{\mu\nu}^{~~~\nu}$ which vanishes by the $h$ equation of motion \eqref{Heom} in the gauge $h=0$, and so the equation is consistently transverse.  
Thus, by solving the wave equation \eqref{xiwaveeqap31} for $\xi_\mu$, we can reach a gauge where
\be
b_{\mu\nu}^{~~~\nu} = 0~,
\label{trbiszero}
\ee
after which there is a residual transverse gauge parameter which satisfies the homogeneous part of \eqref{xiwaveeqap31}
\be
\left(\square + (1+2D)H^2\right)\xi_\mu = 0.\label{residgauge13}
\ee

Plugging the gauge choice~\eqref{trbiszero} into~\eqref{tracefreedivergence} using the gauge \eqref{h0app31gauge} then tells us 
\be
\nabla^\rho b_{\rho\mu\nu}=0. \label{divbisddF}
\ee

Finally, after plugging \eqref{trbiszero}, \eqref{divbisddF} into \eqref{t0curlyG}, we obtain a Klein-Gordon equation for $b_{\mu\nu\rho}$, and the equations of motion reduce to
\be
\big(\square  -3 H^2\big)b_{\mu\nu\rho}=0~,~~~~~~~~~~~b_{\mu\nu}^{~~\nu} = 0~,~~~~~~~~~~~\nabla^\rho b_{\rho\mu\nu} = 0~,~~~~~~~~~~~h=0,
\ee
where there is the residual gauge symmetry
\be  \delta b_{\mu\nu\rho}= b_{\mu\nu\rho} = \nabla_{(\mu}\nabla_{\nu}\xi_{\rho)}+H^2 g_{(\mu\nu}\xi_{\rho)} ,\ \ \ \ \ \left(\square + (1+2D)H^2\right)\xi_\mu = 0~,~~~~~~~~~\nabla_\mu\xi^\mu = 0,
\ee
matching \eqref{spin3t1eomr1}, \eqref{spin3t1eomr2}.

\renewcommand{\em}{}
\bibliographystyle{utphys}
\addcontentsline{toc}{section}{References}
\bibliography{PMdual16}

\end{document}